\documentclass[aps,prl,reprint,superscriptaddress]{revtex4-1}
\usepackage[latin9]{inputenc}
\usepackage{color}
\usepackage{amsmath}
\usepackage{amssymb}
\usepackage{graphicx}
\usepackage[unicode=true,pdfusetitle,
 bookmarks=true,bookmarksnumbered=false,bookmarksopen=false,
 breaklinks=false,pdfborder={0 0 1},backref=false,colorlinks=false]
 {hyperref}
\hypersetup{
 colorlinks,linkcolor=blue,citecolor=blue,urlcolor=blue}

\def\XXint#1#2#3{{\setbox0=\hbox{$#1{#2#3}{\int}$}
     \vcenter{\hbox{$#2#3$}}\kern-.5\wd0}}

\makeatletter

\usepackage{verbatim}
\usepackage{braket}
\usepackage{bbold}
\usepackage{bm}
\renewcommand{\vec}[1]{\boldsymbol{#1}}

\makeatother

\begin{document}

\title{Theory of Spin Injection in Two-dimensional Metals \\
 with Proximity-Induced Spin-Orbit Coupling}
\author{Yu-Hsuan Lin }
\affiliation{Department of Physics, National Tsing Hua University, Hsinchu 30013, Taiwan}
\author{Chunli Huang}
\affiliation{Department of Physics, The University of Texas at Austin, Austin,
Texas 78712,USA}
\affiliation{ National Center for Theoretical Sciences (NCTS), Hsinchu 30013, Taiwan}
\author{Manuel Offidani}
\affiliation{University of York, Department of Physics, YO10 5DD, York, United
Kingdom}
\author{Aires Ferreira}
\affiliation{University of York, Department of Physics, YO10 5DD, York, United
Kingdom}
\author{Miguel A. Cazalilla}
\affiliation{Department of Physics, National Tsing Hua University,  Hsinchu 30013, Taiwan}
\affiliation{ National Center for Theoretical Sciences (NCTS), Hsinchu 30013, Taiwan}
\affiliation{Donostia International Physics Center (DIPC), Manuel de Lardizabal, 4. 20018 Donostia-San Sebastian, Spain}

\begin{abstract}
Spin injection is a powerful experimental probe into a wealth of nonequilibrium spin-dependent phenomena displayed by materials with spin-orbit coupling (SOC). Here, we develop a theory of coupled spin-charge diffusive transport in    two-dimensional spin-valve devices.   The theory describes a realistic proximity-induced SOC  with both spatially uniform and random components of the SOC due to adatoms and imperfections, and applies to the  two dimensional electron gases found in two-dimensional materials and van der Walls heterostructures. The various charge-to-spin conversion mechanisms known to be present in diffusive metals, including  the spin Hall effect and several mechanisms contributing current-induced spin polarization are accounted for. Our analysis shows that the dominant conversion mechanisms can be discerned by analyzing the  nonlocal resistance of the spin-valve for  different polarizations of the injected spins and as a function of the 
applied in-plane magnetic field.  
\end{abstract}
\maketitle

Layer-by-layer assembly of atomically thin crystals
has provided a unique platform to realize emergent phenomena in 
two dimensional electron systems~\citep{Geim_Nature2013}. Examples range from 
secondary Dirac points and Hofstadter's butterfly in Moir$\acute{e}$ superlattices \citep{VDW_Yankowitz_12,VDW_Ponomarenko13,VDW_Dean_13}
to superconductivity in twisted bilayer graphene \citep{VDW_Cao_18,VDW_Yankowitz19}
and long-lived excitons in heterobilayers made from semiconducting
two-dimensional (2D) crystals \citep{VDW_Rivera_15}. 

The engineering of the electronic properties, and in particular, the strength of spin-orbit coupling (SOC) in layered materials
is facilitated by the weak van der Waals bonding that allows the stacking of two-dimensional (2D) crystals with precise
interlayer registry \citep{VDW_Review_Yankowitz_19}. Indeed, several approaches  can be used to enhance and control SOC in the two-dimensional electron gases formed at the atomically thin interfaces  of 2D crystals
and van der Walls heterostructures.  Being essentially surfaces,  
it is possible to decorate them with various kinds of  absorbates that  locally induce/enhance 
the SOC by proximity~\cite{ImpuritySOC_1,
ImpuritySOC_2,ImpuritySOC_3,SHE_Experiment_theta}.  In addition,  
 SOC can be substantially enhanced by placing graphene  
 layers on semiconducting transition metal dichalcogenides (TMD) monolayers \citep{GTMD_Avsar_14,GTMD_Gmitra_16,GTMD_Wang_15,GTMD_Alsharari_18,GTMD_Wang_16a,GTMD_Volkl_16b,GTMD_Yang_17,GTMD_Wakamura_18a,GTMD_Omar_18b,GTMD_Optical_Avsar_ACSNano2017,GTMD_SRTA_Ghiasi_17,GTMD_SRTA_Benitez18,GTMD_SRTA_Offidani_18,GTMD_Review_Garcia_18,Luo_NanoLett2017,Offidani_PRL2017}.
A sizeable and controllable SOC in graphene and other 2D metals
provides a rich arena for the study spin transport phenomena that is not accessible in other, more conventional 2D
metals
such as resonantly-enchanced skew scattering from spin-active  impurities~\citep{Ferreira_14,Huang_PRB2016,Milletari_Ferreira_PRB2016,Milletari_Ferreira_PRB2016_Rapid}, or
spin-transparent impurities in graphene  with noncollinear spin texture \citep{GProximitySOC_Milletari_PRL2017},
as well as anisotropic-spin precession scattering 
from impurities that induce Rashba-like SOC by proximity~\citep{Huang_PRB2016}.

Previous studies have modeled  proximity-induced SOC in heterostructures made of graphene on TMDs by 
treating the interfacial coupling as a perturbation to the band structure that
is compatible with the lattice symmetries of pristine graphene~\citep{GProximitySOC_Milletari_PRL2017,Huang_Milletari_Cazalilla,
Offidani_PRL2017,Offidani_MDPI2018}. This minimal model treats the proximity-induced SOC as ``intrinsic''
and reproduces accurately the spin splitting and $\mathbf{k}$-dependent
spin polarization of low-energy states from first-principles calculations
\citep{GTMD_Wang_15,GTMD_Gmitra_16,GTMD_Alsharari_18}. Thus, it may
be regarded as an accurate description of ultra-clean  heterostructures,
where conduction states lie within the band gap of the substrate and
are therefore only weakly affected by interfacial SOC. However, a realistic model should also contain a spatially fluctuating SOC component that describes, for example, structural inhomogeneities between the two materials. 
Moreover, random SOC-active impurities \cite{ImpuritySOC_1,ImpuritySOC_2} are inevitable even in the cleanest samples~\citep{GTMD_Wang_16a}. Owing to the Dirac nature of charge carriers in some 2D materials, localized spin-orbit potentials can lead to sharp scattering resonances and thus enhanced skew scattering \cite{Ferreira_14}. The kinetic theory formulated in Ref.~\citep{Huang_PRB2016,Huang_PRL2017}
describes spin-coherent transport in single-layer graphene containing a dilute
ensemble of SOC-active impurities. Notably, current-induced spin polarization (CISP) can arise purely from random SOC ~\citep{Huang_PRB2016,Huang_PRL2017}: In addition to extrinsic version of the  Edelstein effect (EE) \cite{Refs_ISGE},
a different (direct) mechanism for spin-charge conversion mechanism was  also found in Ref.~\cite{Huang_PRB2016}. Termed anisotropic spin precession scattering~\citep{Huang_PRB2016,Huang_PRL2017}, it is a direct mangeto-electric
effect (DMC)  which yields an additional contribution to the CISP. 

In this work, we  study spin injection in spin-valve devices made from 2D metals with SOC induced by proximity. 
In such devices, we have found that the polarization of the injected spins determines the dominant spin-to-charge conversion mechanism at  distances $\sim l_s$ where $l_s$ is the spin-diffusion length. Thus, it is possible to ascertain which mechanism  yields the dominant contribution to the nonlocal resistance of the device by controlling the polarization of the injected spins or by analyzing the dependence of the nonlocal resistance with an in-plane magnetic field. The two mechanisms  that can contribute to the nonlocal resistance are  either  the inverse SHE or the inverse CISP (also known as spin-Galvanic effect, SGE). Both mechanisms are the Onsager reciprocal of the SHE and the CISP. However, for sake of simplicity, below we shall refer to them as SHE and CISP.

 Furthermore, below we also provide a  microscopic derivation from kinetic theory of the spin diffusion equations describing diffusive transport in 2D metals where the proximity-induced SOC contains randomly fluctuating components. To this end, we consider two distinct physical scenarios. First, we consider a model of random SOC induced by impurities. The single-impurity potential is treated by means of the T-matrix approach, which allows us to  capture resonant-scattering effects. In a second scenario, the proximity-induced SOC potential consists of a uniform (``intrinsic?  component and a random component, which is treated in the gaussian (i.e. ``white noise'') approximation. 
 We show that  these two scenarios lead to the same set of drift-diffusion equations, albeit with different values for the transport and spin-charge conversion coefficients. Thus, we expect  this set of equations will apply to a fairly broad class of 2D diffusive metals  with proximity-induced SOC.

 The remainder of the manuscript is organized as follows. In Sec.~\ref{sec:DDE},
we present the set of drift-diffusion equations thatand briefly discuss how they compare to those derived in previous 
works.  In Sec.~\ref{sec:spin-valve},
the equations are applied to a non-local spin valve device and the smoking-gun signatures of the charge-to-spin conversion are discussed.  Sections~\ref{sec:boltzmann} and \ref{sec:su2} are concerned with the microscopic derivation of the spin-charge coefficients for uniform proximity-induced SOC (Sec.~\ref{sec:boltzmann}) and random SOC (Sec.~\ref{sec:su2}).

\section{Coupled spin-charge diffusion equations}\label{sec:DDE}

In the diffusive regime where the elastic mean free
path $\ell$ is much larger than the Fermi wavelength $k_{F}^{-1}$, the  coupled spin-charge dynamics is described by the
following set of equations (henceforth summation over repeated indices is implied unless otherwise stated):
\begin{equation}
\partial_{t}\rho+\partial_{i}J_{i}=0,\label{eq:c-cont}
\end{equation}
\begin{equation}
[\nabla_{t}s]^{a}+\big[\nabla_{i}\mathcal{J}_{i}\big]^{a}=- \Gamma^{ab}_s s^{b} +\kappa_{i}^{a}J_{i},\label{eq:s-cont}
\end{equation}
\begin{equation}
J_{i}=-D\left(\partial_{i}\rho+\kappa_{a}^{i}s^{a}\right)+\gamma_{ij}^{a}\mathcal{J}_{j}^{a},\label{eq:c-consti}
\end{equation}
\begin{equation}
\mathcal{J}_{i}^{a}=-D\left[\nabla_{i}s\right]^{a}+\gamma_{ij}^{a}J_{j},\label{eq:s-consti}
\end{equation}
where we have used the following notation: 
\begin{align}
[\nabla_{i}O]^{a} & =\partial_{i}O^{a}-\epsilon^{abc} A_{i}^{b}O^{c},\label{eq:cov-der}\\{}
[\nabla_{t}O]^{a} & =\partial_{t}O^{a}+\epsilon^{abc} A_{0}^{b}O^{c}.
\end{align}
Eqs.~(\ref{eq:c-cont}) and (\ref{eq:s-cont}) are the continuity
equations for the charge carrier density ($\rho$) and electron's spin
density ($s^{a}$, where $a \in \{x,y,z\}$), respectively. 
$\Gamma_{s}^{ab}$ are the (anisotropic)  relaxation rates for the spin;  $J_{i}$ and $\mathcal{J}_{i}^{a}$
are  the charge  and spin current densities, respectively,  and  $i\in \{x,y\}$. 
Eqs.~(\ref{eq:c-consti}) and (\ref{eq:s-consti}) are the
generalized constitutive relations for the local charge and spin observables; $D$
is the diffusion constant, which we have assumed to be the same for
charge and spin (relaxing this assumption only affects our results quantitatively at the cost of introducing
additional complexity). 
The coupling between charge current ($J_{i}$), spin current ($\mathcal{J}_{i}^{a}$)
and spin density ($s^{a}$) is described by two sets of spin-charge conversion
rates: $\gamma_{ij}^{a}$ controls the magnitude spin Hall effect (SHE), and
$\kappa_{i}^{a} = -\kappa^{i}_a$ controls the magnitude of the direct magneto-electric (DMC) coupling~\cite{Huang_PRL2017}, a contribution to current-induced spin polarization (CISP) 
additional to the Edelstein effect (EE)~\cite{Refs_ISGE}. In addition, the coupling between $\mathcal{J}_{i}^{a}$
to $s^{a}$ is hidden in the covariant derivative defined in Eq.~\eqref{eq:cov-der}.
In this equation, $A_{i}^{a}$ describes the coupling to the uniform
component of the Rashba-type SOC and $A_{0}^{b}=g \mu_{L} \mathcal{H}^{b}$ describes
the Zeeman coupling. The discussion of spin-swapping \citep{Lifshits_PRL2009}
term in Eq.~\eqref{eq:s-consti} is relegated to Sec.~\ref{sec:boltzmann}
since they are not directly related to spin-charge current, and we
treat $A_{i}^{b}$, $\gamma_{ij}^{a}$, $\kappa_{i}^{a}$ in Eqs.~\eqref{eq:c-consti} to~\eqref{eq:cov-der} phenomenologically
since they are model-dependent as shown in Section~\ref{sec:boltzmann} and ~\ref{sec:su2}.

It is useful to compare the above set of equations, ~\eqref{eq:c-cont}  to \eqref{eq:s-consti},
with those derived in previous work. A similar set of coupled spin-charge diffusion equations were derived for 2D electron gases by means of the Kelydsh formalism with SOC treated as a non-Abelian (SU(2)) gauge field in Refs.~\citep{Shen_PRB2014,Shen_PRL2014}. However, in addition to the spin-charge conversion mechanisms 
described therein, Eqs.~\eqref{eq:s-cont}  and \eqref{eq:c-consti} also account for the DMC mechanism.
 The latter describes a (direct)  coupling between
the charge current, $J_{i}$, and the spin polarization, $s^{a}$, and it is parametrized by  the coefficients 
$\kappa_{i}^{a} = -\kappa^{i}_a$. We shall show in Secs.~\ref{sec:boltzmann} and ~\ref{sec:su2} that the
DMC can emerge from the scattering of the carriers with the spatially random components of the SOC, and
more specifically, from a non-vanishing correlation between in-plane and out-of-plane electric fields at the interface.

   In Refs.~\citep{Burkov_PRB2004,Burkov_PRL2010}, the
spin diffusion equations were derived from the density-density
response function. This approach is  well suited in the strong SOC regime where the intrinsic 
SOC is comparable to the Fermi energy, as in the case of surface states of 
3D topological insulator~\citep{Burkov_PRL2010}. Such strong SOC regime, 
strictly speaking, lies outside the applicability 
of the  microscopic models discussed in Sec.~\ref{sec:boltzmann} and \ref{sec:su2} 
and used to derive Eqs.~\eqref{eq:c-cont}  to \eqref{eq:s-consti}. Nevertheless,
on phenomenological grounds,  it is worth exploring how such regime can be 
described starting from the above set of equations.
In the strong SOC regime, the spin current is not a hydrodynamic mode 
of the system and the only  relevant spin-charge conversion rate 
corresponds to  $\kappa_{i}^{a}$ in Eq.~\eqref{eq:s-cont} for the DMC.  Thus, upon setting
$\gamma_{ij}^{a}=0$  in Eq.\eqref{eq:c-consti}, we recover Eq.~5
of Ref.~\citep{Burkov_PRL2010} with $\kappa_{i}^{a}=\ell^{-1} \epsilon^{a}_i$, $\ell = v_F \tau$ ($\tau$) 
being the  elastic mean-free path (elastic scattering time). Finally, we note that a
similar set of  equations has  been obtained for superconductors
within the quasi-classical approximation in 
Refs.~\citep{Bergeret_PRB2014,Bergeret_PRB2016,Huang_PRB2018}.
The latter are complicated by the fact that quasi-particle spectral
weights are no longer peaked on the Fermi surface and in general are
altered by the nonequilibrium dynamics. However, in the normal state,
they can be brought to the form of Eqs.~\eqref{eq:c-cont}-\eqref{eq:s-consti}.
\begin{figure}
\centering{}\includegraphics[scale=0.4]{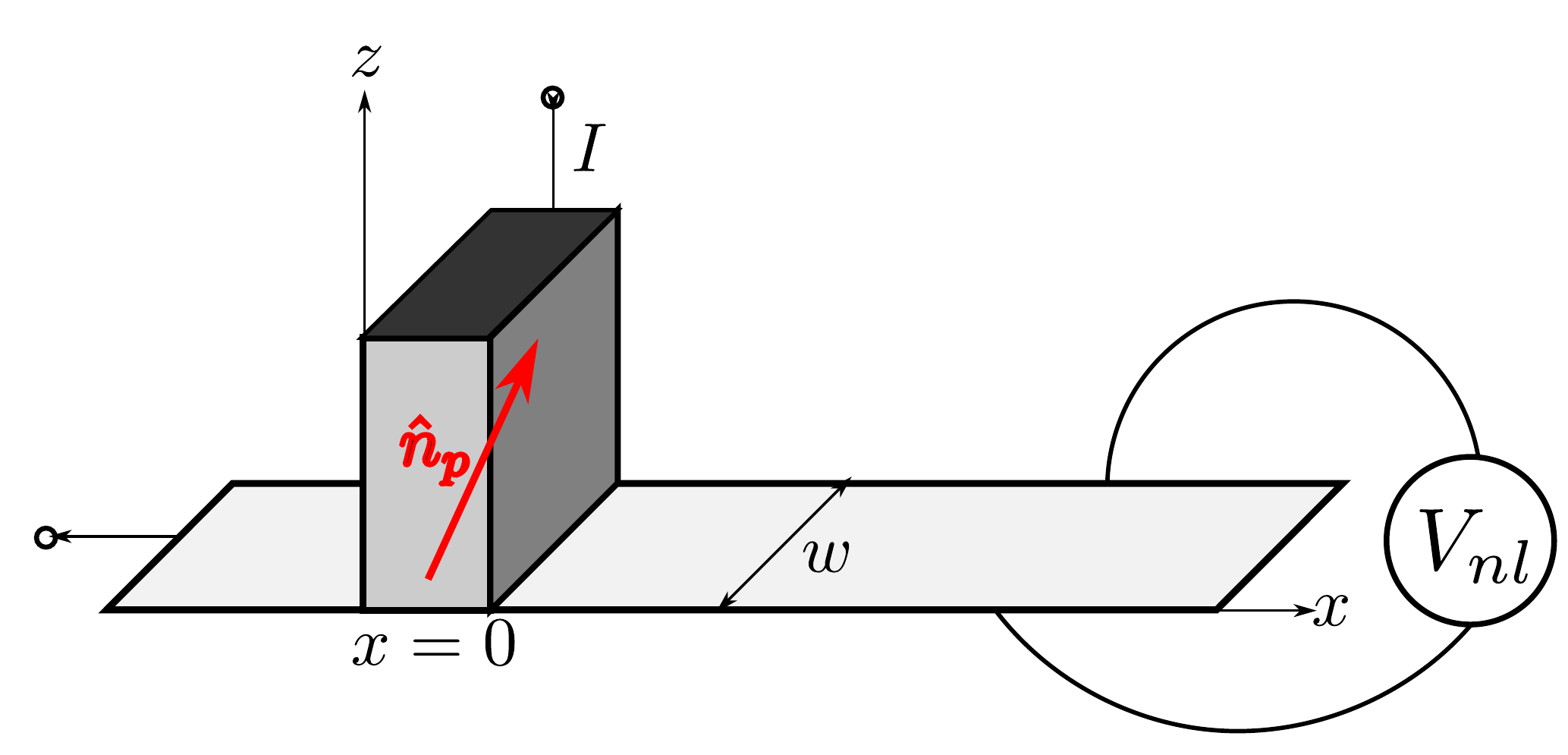} \caption{Illustration of the nonlocal transport device considered
in this work. The external magnetic $\vec{\mathcal{H}}$ field is applied along the $y$ axis, on the plane of
the device. \label{fig: device pic}}
\end{figure}

\section{Spin-Valve}\label{sec:spin-valve}

In this section, our goal is to describe the properties of the nonlocal resistance in a lateral spin-valve device of the type employed to measure the inverse spin Hall effect in the seminal experiments by Valenzuela and Tinkham~\cite{sergio-tinkham}, see Fig.~~\ref{fig: device pic} for an illustration of the device.

We shall be concerned with 2D metals that are isotropic in the long wavelegth limit, but, due to presence of 
a substrate or absorbates,  have broken mirror reflection symmetry about the 2D plane. This includes van der Waals heterostructures, such as graphene on TMD~\citep{Cysne18}. From these symmetry considerations, the conversion rates describing the SHE and DMC are given by: 
\begin{equation}
\gamma_{ij}^{a}=\theta_{\textrm{sH}}\epsilon_{ij}\delta^{az}\qquad\kappa_{i}^{a}=l_{\textrm{DMC}}^{-1}\epsilon_{i}^{\,\, a},
\end{equation}
where $\theta_{\text{sH}}$ is the spin Hall angle and $l_{\textrm{DMC}}$
 is a parameter with units of length that determines the conversion efficiency of the DMC  ($\epsilon_x^{\, \, y} = \epsilon^{x}_{\,\,y} =  \epsilon_{xy} = - \epsilon_y^{\, \, x} = -\epsilon^y_{\,\,x}  = -\epsilon_{yx} = 1$ is the fully anti-symmetric 2D tensor). In addition,
\begin{equation}
A_{i}^{a}=l_{R}^{-1}\epsilon^{a}_{\,\,i}
\end{equation}
where the parameter $l_R$ has units of length and parametrizes the strength of the inversion-symmetry breaking Rashba SOC (cf. Sec.\ref{sec:boltzmann} and \ref{sec:su2}). In order to reduce the number of parameters 
in the model calculation below, we shall assume that the spin relaxation time to be isotropic: 
$\Gamma_{s}^{ab}=\delta^{ab}\tau^{-1}_{s}$ (i.e. it is the same for the in-plane and out of
plane spin components).   These assumptions will allow us to derive simple analytical expressions
for the nonlocal resistance of the device ((see Ref.~\ref{Lin2019} for a discussion of the corrections to the
nonlocal  transport introduced by spin lifetime anisotropy). 

 In what follows, we shall work in the limit where SOC is weak compared to the Fermi energy of the electron gas.
 Therefore,  the spin diffusion length $l_s = \sqrt{D\tau_{s}}\gg \ell$. In addition, the  
 dimensionless spin-charge conversion ratios $\theta_{\textrm{sH}}$,   $l_s/l_{\textrm{DMC}}$,
 and  $l_s/l_{R}$ will be assumed  to be small (compared to unity) and therefore
contributions of quadratic order in these coefficients can be safely neglected.
Under such conditions,  the build-up of a non-local voltage in the lateral spin valve 
(Fig.~\ref{fig: device pic}) can be regarded as the result of a three-stage process. First, a finite spin density,
$\vec{s}(x=0)$, is injected by driving a current  $I$ through the ferromagnetic metal
contact. Second, the injected spin polarization $\vec{s}(x=0)$  diffuses away from the injection
point according to Eq.~\eqref{eq:s-cont}. And finally,  at a distance $x$  from
the injector,  $\vec{s}(x)$  generates a transverse electric current via
Eq.~\eqref{eq:c-consti} and leads to the appearance of  a finite nonlocal voltage, $V_{nl}(x)$ 
The measured nonlocal resistance, $R_{nl}(x)$ is the ratio $V_{nl}(x)/I$.  Notice that, for large SOC, 
this three stages  are not independent and one has
to solve Eqs.~\eqref{eq:c-cont} to \eqref{eq:s-consti} self-consistently,
see e.g.~Ref.~\citep{Zhang_2DMat2017}. In the following, we shall
describe the three stages in detail.

\subsection{Spin-injection}

For a ferromagnetic metal contact whose dimensions are much smaller than
the spin diffusion length  ($l_{s}$) in the 2D material,
the injected spin density can be described by a single vector $\vec{s}(x=0)$ whose direction
and magnitude depends on the details of the contact. From the conservation
of charge and spin current at the contact, the following boundary
conditions are obtained~\citep{takahashi_PRB_2003}: 
\begin{align}
J_{\mathrm{F}}(z=0) & =J(x=0),\label{eq:bc1}
\end{align}
\begin{align}
\mathcal{J}_{\mathrm{F}}(z=0) & =\hat{\boldsymbol{n}}_{p}\cdot\left[\boldsymbol{\mathcal{J}}_{x}\left(x=0^{+}\right)-\boldsymbol{\mathcal{J}}_{x}\left(x=0^{-}\right)\right].\label{eq:bc2}
\end{align}
Here, $J_{\mathrm{F}}$ and $\mathcal{J}_{\mathrm{F}}$
are, the charge and the spin current densities flowing
into the 2D metal, respectively,
and $\hat{\vec{n}}_{p}=\sin\theta_{p}\cos\varphi_{p}\hat{x}+\sin\theta_{p}\sin\varphi_{p}\hat{y}+\cos\theta_{p}\hat{z}$ is
 the polarization direction of the injected spins near the contact. 
Eqs.~\eqref{eq:bc1} and~\eqref{eq:bc2} assume
that the contact does not trap charge or accumulate any spin torque.
In this situation, the spin polarization of the injected carriers is parallel
to the ferromagnet magnetization. Thus, as we show below, the magnitude of the spin density
depends on the applied current $I$ and the contact conductance.

At the contact position (i.e. $x = 0$), the terms proportional to the gradient of the charge
and spin densities in the constitutive relations (cf. Eq.\eqref{eq:c-consti}
and \eqref{eq:s-consti}) dominate. Thus, we can approximate 
\begin{align}
J(x=0) & \approx-D\frac{d\rho(x)}{dx}\bigg|_{x=0},\\
\hat{\boldsymbol{n}}_{p}\cdot\boldsymbol{\mathcal{J}}_{x}(x=0^{\pm}) & \approx-D\frac{d\,(\vec{s}(x)\cdot\vec{\hat{n}}_{p})}{dx}\bigg|_{x=0^{\pm}}.
\end{align}
\subsection{Spin diffusion away from injection}

 Next, we derive the spin diffusion (Bloch) equation from the set of
drift-diffusion equations introduced in Sec.~\ref{sec:DDE} by eliminating the
charge and spin-currents. In addition, 
we shall assume that the spin channel in the 2D metal has a 
large length-to-width ratio $L/w\gg1$ and also $w\ll l_{s}$, so that the spin relaxation along the transverse
direction is suppressed. Within this one-dimensional channel
 approximation,  the resulting spin diffusion equation can be written as follows:
\begin{equation}
\vec{\mathcal{\bar{D}}}\cdot\vec{s}(x)+\omega_{L}\left(\vec{\hat{n}}_{H}\times\vec{s}\left(x\right)\right)=0\label{eq: Diffusion equation}
\end{equation}
where
\begin{equation}
\boldsymbol{\mathcal{\bar{D}}}=D\left(\begin{array}{ccc}
\partial_{x}^{2}-l_{s}^{-2} & 0 & 2l_{R}^{-1}\partial_{x}\\
0 & \partial_{x}^{2}-l_{s}^{-2} & 0\\
-2l_{R}^{-1}\partial_{x} & 0 & \partial_{x}^{2}-l_{s}^{-2}
\end{array}\right)\label{eq: Diffusion matrix isotropic}
\end{equation}
and $\omega_{L}=g\mu_{L}\left|\vec{\mathcal{H}}\right|/\hbar$ is the
Larmor  frequency induced by the magnetic field $\vec{\mathcal{H}} = \left|\vec{\mathcal{H}}\right| \vec{\hat{y}}$, and
$\vec{\hat{n}}_H = \vec{\hat{y}}$. 


The general solution to Eq.~\eqref{eq: Diffusion equation} can be written as follows:
\begin{align}
s^{x}(x) & =s^{x}(0)\mathrm{Re}\,z(x)-s^{z}(0)\mathrm{Im}\,z(x)\\
s^{z}(x) & =s^{z}(0)\mathrm{Re}\,z(x)+s^{x}(0)\mathrm{Im}\,z(x). \label{eq:soldiff}
\end{align}
The $s^{y}(x)$ component decouples from the others and does not contribute to the
spin-charge conversion processes (its behavior is discussed in Appendix~\ref{app:spin-valve}).
The function $z(x)$ characterizes the oscillatory decay of
the two spin components and reads: 
\begin{equation}
z(x)=\mathrm{exp}\left(-\kappa|x|+i\frac{x}{l_{R}}\right),
\end{equation}
where $\kappa=\sqrt{l_{s}^{-2}-l_{R}^{-2}+i\omega_{L}D^{-1}}$
and the two constants, $s_{x}(0)$ and $s_{z}(0)$ are obtained by matching the solution with
the boundary conditions, Eqs.~\eqref{eq:bc1} and \eqref{eq:bc2}. The
calculation of $s_{x}(0)$ and $s_{z}(0)$ is described in Appendix~\ref{app:spin-valve}.
Here it suffices to know that the result depends on the
injected current $I$ and the conductance of the junction between the
ferromagnetic metal contact and the 2D material.

\subsection{Spin-charge conversion and nonlocal voltage}

\begin{figure*}
\centering{}\includegraphics[width=2\columnwidth]{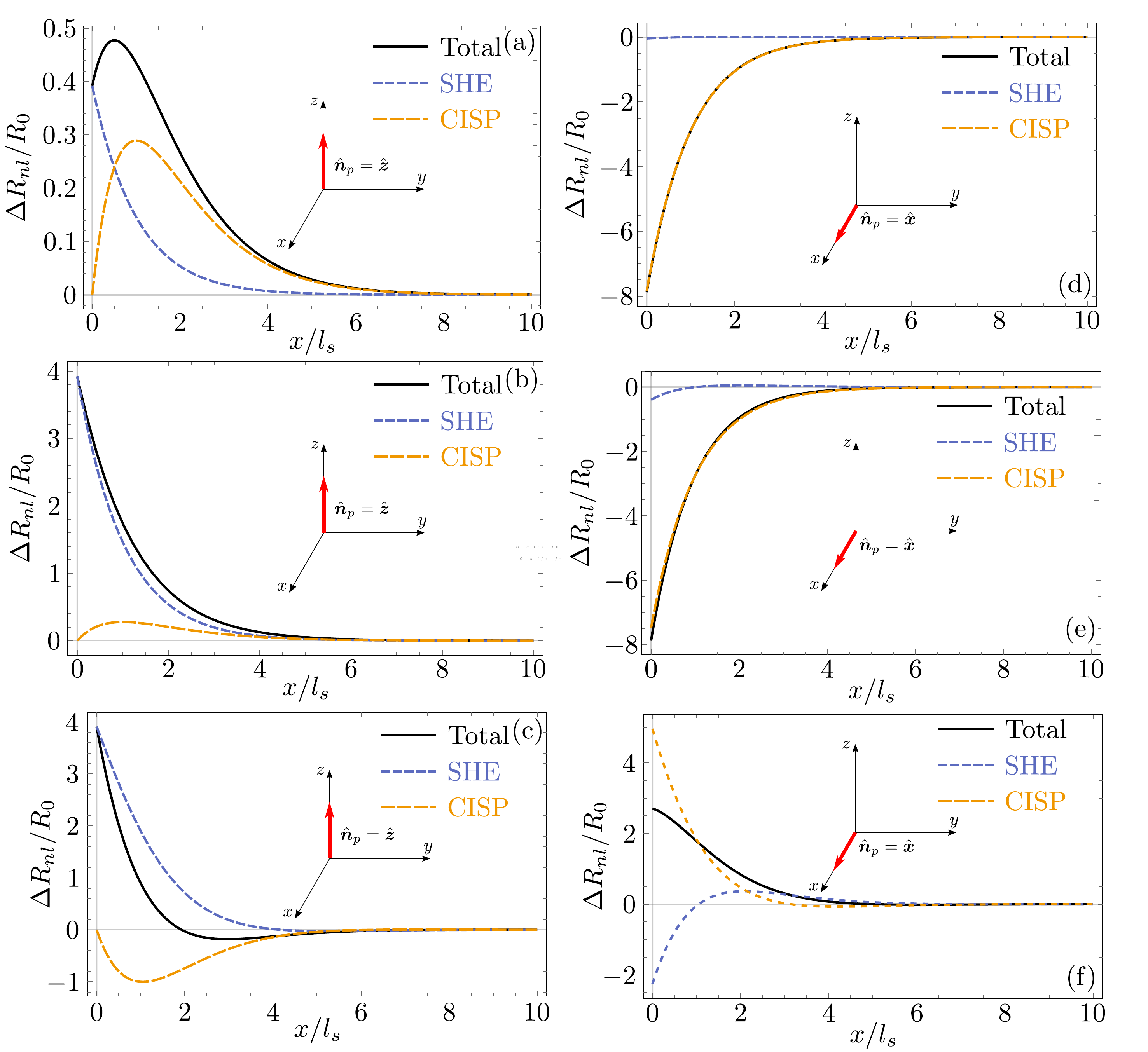}
 \caption{Nonlocal resistance $R_{nl}(x)$ versus distance from the spin injection contact ($x$). In panels a,b, and c (d,e, and ef) the polarization of the injected spins is  perpendicular (parallel) to the plane  of the 2D electron gas. The results depend on three  spin-charge conversion coefficients, namely the spin-Hall angle $\theta_{\mathrm{sH}}$, a length scale associated with the spin precession induced by the Rashba SOC, $l_R$ and a length scale associated with a direct magneto-electric coupling, $l_{\mathrm{DMC}}$. For each panel, we have chosen the following experimentally relevant  values: $l_{s}=10^{-6}$m \citep{SHE_Experiment_theta}; $\theta_{\textrm{sH}}=-0.01$,  $l_{R}=2l_{\textrm{DMC}}=10l_{s}$ in (a) and (d);
 $\theta_{\textrm{sH}}=-0.1$,  $l_{R}=2l_{\textrm{DMC}}=10l_{s}$ in (b) and (e);
 $\theta_{\textrm{sH}}=-0.1$,  $l_{R}=-0.12l_{\textrm{DMC}}=2l_{s}$ in (c) and (f);
 $P_{J}=0.4$ \citep{SHE_Experiment_PJ}, $P_{F}=0.73$ \citep{SHE_Experiment_PF}, $G_{N}/G_{F}=0.01$ \citep{SHE_Experiment_R_ratio}, and $G/G_{F}=5\times10^{-4}$. }
 \label{fig:Rnl-x decomposition}
\end{figure*}

\begin{figure*}
\includegraphics[width=2\columnwidth]{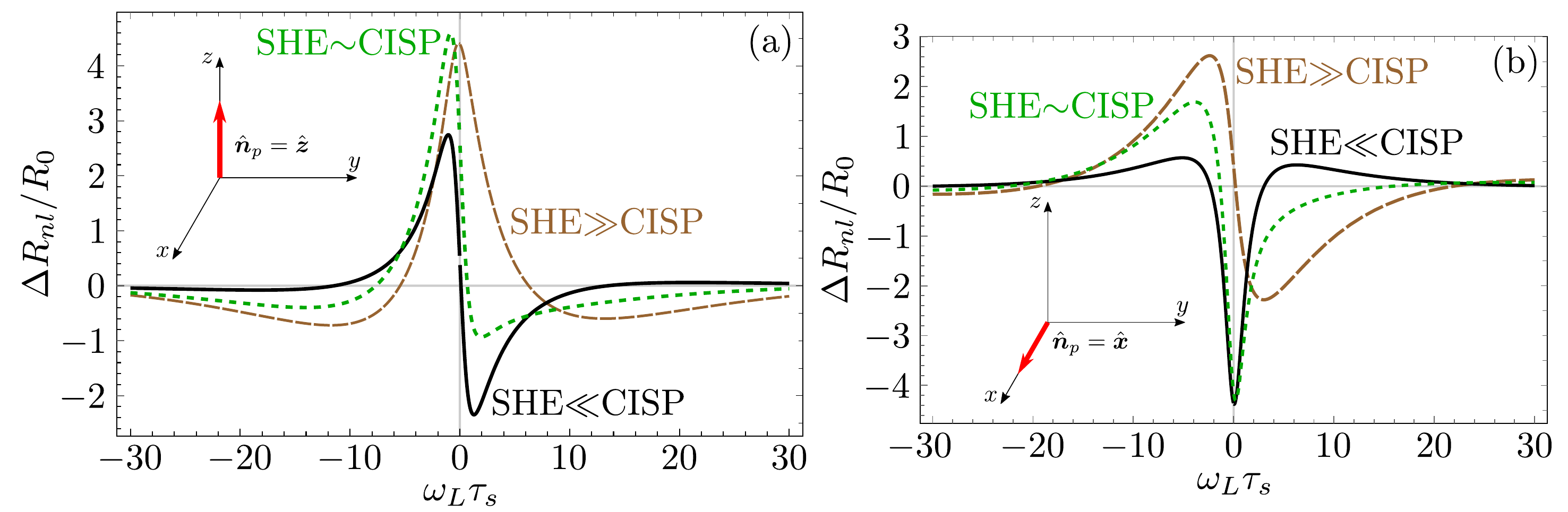}
\label{fig:Nonlocal resistance versus external in-plane magnetic field}
\caption{
Nonlocal resistance versus magnetic field (measured in units of the Larmor frequency times the spin relaxation time,
i.e. $\omega_L \tau_s$) 
at $x=l_{s}$. We take $l_{s}/l_{R}=0.1$ for all curves.
The parameters for the solid black curve are $\theta_{\textrm{sH}}=-10^{-3}$, and
$l_{s}/l_{\textrm{DMC}}=0.2$. The parameters for the dashed (brown)
curve are $\theta_{\textrm{sH}}=-0.2$, and $l_{s}/l_{\textrm{DMC}}=2\times10^{-3}$.
The parameters for the dashed (green) curve are $\theta_{\textrm{sH}}=-0.1$, and
$l_{s}/l_{\textrm{DMC}}=0.2$.\label{fig:Nonlocal resistance versus magnetic field}}
\end{figure*}

Next, we use the solution of the spin Bloch equation to obtain the charge current flowing
along the $y$-direction, $J_{y}(x)$. This transverse electric current
generates a voltage drop $V_{\textrm{nl}}(x)$. The nonlocal resistance is thus defined 
by the expression:
\begin{equation}
R_{\textrm{nl}}(x)=\frac{V_{\textrm{nl}}(x)}{I}=\frac{wJ_{y}\left(x\right)}{I\sigma_{N}},
\end{equation}
where $\sigma_{N}$ is the electric conductivity of the device and $w$ is the channel width. The solution of the spin diffusion equations contains three distinct contributions to the nonlocal signal:
\begin{align}
R_{\textrm{nl},\textrm{sH}}\left(x\right)=\frac{wD}{I\sigma_{N}}\theta_{\textrm{sH}}\partial_{x}s^{z}\label{eq:Rnl_sH}
\end{align}
\begin{align}
R_{\textrm{nl},\textrm{EE}}\left(x\right) & =-\frac{wD}{I\sigma_{N}}\theta_{\textrm{sH}}l_{R}^{-1}s^{x}\label{eq:Rnl_Edelstein}
\end{align}
\begin{align}
R_{\textrm{nl},\textrm{DMC}}\left(x\right) & =-\frac{wD}{I\sigma_{N}}l_{\textrm{DMC}}^{-1}s^{x}\label{eq:Rnl_DMC}
\end{align}
Experimentally, $R_{\textrm{nl},\textrm{EE}}(x)$ and $R_{\textrm{nl},\textrm{DMC}}(x)$ cannot be distinguished and
therefore  we shall combine them into one single contribution to $R_{\textrm{nl}}(x)$ arising from the current-induced spin polarization (CISP) mechanisms:
\begin{align}
R_{\textrm{nl},\textrm{CISP}}\left(x\right)&=R_{\textrm{nl},\textrm{EE}}\left(x\right)+R_{\textrm{nl},\textrm{DMC}}\left(x\right)\label{eq:Rnl_CISP}
\end{align}
In realistic spin-valve measurements, there is always some level of background noise, which masks the pure spin contribution to the nonlocal resistance \citep {sergio-tinkham}. The background signal can be eliminated by subtracting the nonlocal resistances between parallel and anti-parallel configurations (see Appendix~\ref{app:spin-valve} for
details): 
\begin{align}
\Delta R_{\textrm{nl}}\left(x\right) & =R_{\textrm{nl}}\left(x\right)\bigg|_{\vec{\hat{n}}_{p}}-R_{\textrm{nl}}\left(x\right)\bigg|_{-\vec{\hat{n}}_{p}}\nonumber \\
 & =R_{0}C_{\textrm{inj}}e^{-\tilde{q}\cos\theta_{L}x}f(\vec{\hat{n}}_{p},\omega_{L})\label{eq: Non-local resistance}
\end{align}
In the above expression,
\begin{align}
\tilde{q} = |\kappa|  = \frac{1}{l_s} \left[\left(1-l_{s}^{2}/l_{R}^{2}\right)^{2}+\left(\omega_{L}\tau_{s}\right)^{2}\right]^{1/4},
\end{align}
 is the characteristic wave number associated with spatial variation of the nonlocal resistance,
 $\theta_{L}  = \tfrac{1}{2} \tan^{-1}\left[\omega_{L}\tau_{s}/\left(1-l_{s}^{2}/l_{R}^{2}\right)\right]\approx
  \tfrac{1}{2} \tan^{-1}\left( \omega_{L}\tau_{s}\right)$, 
 and $R_{0}=(w/l_{s}) G_{F}$, where $G_F$ is the conductance of the ferromagnetic metal. The dimensionless parameter $C_{\textrm{inj}}$ characterizes
the properties of the junction between the ferromagnet and the 2D
material. Typically, the conductance of the normal metal is much smaller
than the ferromagnet $G_{N}/G_{F}\sim10^{-2}$ (tunneling limit).
Thus, in this regime where $G_{N}\gg G_F$, the injection spin efficiency becomes:
\begin{equation}
C_{\textrm{inj}}\simeq\frac{P_{J}G_{F}}{G_{N}\tilde{q}l_{s}}.
\end{equation}
On the other hand, in the transparent limit where $G\gg G_{F}$, 
\begin{equation}
C_{\textrm{inj}}\simeq\frac{2P_{F}}{1-P_{F}^{2}}\frac{1}{\cos\theta_{L}+\left(\tilde{q}l_{s}-\cos\theta_{L}\right)\sin^{2}\theta_{p}\sin^{2}\varphi_{p}}.
\end{equation}
The dimensionless function $f(\vec{\hat{n}}_{p},\omega_{L})$ in Eq.~(\ref{eq: Non-local resistance})
describes the interplay between different spin-charge conversion effects, the Larmor precession, and the
quantization axis (magnetization direction) of the ferromagnet described by $\vec{n}_{p}$. Its full form is given in Eq.~\eqref{eq:f full form} in  Appendix~\ref{app:spin-valve}.

Let us first discuss the main features of the nonlocal resistance in the absence of magnetic field, i.e.~$f(\vec{\hat{n}}_{p},\omega_{L}=0)$. It takes the following  form for $\vec{n}_{p}$ along the
in the $x$ and $z$ axes, respectively:
\begin{align}
f(\vec{\hat{z}},0)= & -\theta_{\textrm{sH}}\tilde{q}l_{s}\cos\left(\frac{x}{l_{R}}\right)+\frac{l_{s}}{l_{\textrm{DMC}}}\sin\left(\frac{x}{l_{R}}\right)\label{eq:f1}\\
f(\vec{\hat{x}},0)= & -\theta_{\textrm{sH}}\tilde{q}l_{s}\sin\left(\frac{x}{l_{R}}\right)-\frac{l_{s}}{l_{\textrm{DMC}}}\cos\left(\frac{x}{l_{R}}\right). \label{eq:f2}
\end{align}
From the above expressions, it can be seen that, up to an exponential decay factor
 (cf. Eq.~\ref{eq: Non-local resistance}), for $x \ll l_{R}$, the nonlocal resistance 
$\Delta R_{\textrm{nl}}(x)\sim \theta_{\textrm{sH}}$
for $\vec{\hat{n}}_{p}=\vec{\hat{z}}$, whereas $\Delta R_{nl}\sim l_{s}/l_{\textrm{DMC}}$
for $\vec{\hat{n}}_{p}=\vec{\hat{x}}$.  
Thus, at distances much smaller than the typical distance
for precession under the Rashba field, $l_R$, the non-local resistance is approximately proportional  
to the spin Hall angle $\theta_{\mathrm{sH}}$  when the injected spins are polarized out 
of the plane of the device, i.e. for $\vec{\hat{n}}_{p}=\vec{\hat{z}}$. On the other hand, 
the nonlocal resistance is approximately proportional to the ratio $l_s/l_{\mathrm{DMC}}$ 
when the injected spins lie on the plane 
of the device, i.e. for $\vec{\hat{n}}_{p}=\vec{\hat{x}}$. The full spatial
dependence  of  $\Delta R_{\textrm{nl}}(x)$
 for zero magnetic field  is shown in Fig.~\ref{fig:Rnl-x decomposition}. The left
panels correspond to  out-of-plane polarization ($\vec{\hat{n}}_{p}=\vec{\hat{z}}$)
whereas the right panels correspond to in-plane polarization  ($\vec{\hat{n}}_{p}=\vec{\hat{x}}$).

  The above observations  concerning the behavior of $\Delta R_{\textrm{nl}}(x)$ 
at short distances point to  possibility of  measuring the spin-charge conversion coefficients $\theta_{\mathrm{sH}}$ and  $l_s/l_{\mathrm{DMC}}$ or at least experimentally discerning the dominant spin-charge conversion mechanism
in a device.  Theoretically,   these coefficients (together with $l_s/l_R$) 
depend on the microscopic details of the model (see Secs.~\ref{sec:boltzmann} and \ref{sec:su2}) and
we have treated them phenomenologically.  Thus,  in  Fig.~\ref{fig:Rnl-x decomposition}, we have plotted
$R_{\mathrm{nl}}(x)$ for a wide range of choices of $\theta_{\mathrm{sH}}$, $l_s/l_R$, and $l_s/l_{\mathrm{DMC}}$. 
The two contributions to $\Delta R_{nl}(x)$ arising from the SHE and CISP mechanisms 
are also displayed in  Fig.~\ref{fig:Rnl-x decomposition} (dashed lines).
Notice that the SHE  is   dominant   for  $\vec{\hat{n}}_{p}=\vec{\hat{z}}$ and
CISP  is  dominant for  $\vec{\hat{n}}_{p}=\vec{\hat{x}}$, as noted above.
However, this does not mean that  the CISP (SHE) contribution  is negligible in the former (latter)
case.  Indeed, a word of caution is necessary since the SHE contribution does not only
correspond to the first term ($\propto \tilde{q}\theta_{\mathrm{sH}}$) 
in the right-hand side of Eqs.~\eqref{eq:f1} and~\eqref{eq:f2}). By the same token,
the second term in Eqs.~\eqref{eq:f1} and~\eqref{eq:f2}) does not exactly correspond to the CISP contribution:
It arises from the DMC contribution. Indeed,  there is an additional term in the expression for the SHE contribution  
which is equal in magnitude but opposite in sign to the EE contribution to CISP ($\propto \theta_{\mathrm{sH}}l_s/l_R$). 
This explains why in the bottom right panel the contribution from SHE  takes
a non-zero value at $x = 0$ despite that the injected spins point along the $x$-axis. 
Indeed, $R_{\textrm{nl},\mathrm{sH}}(x=0) \sim \partial_x s^x(x = 0)  = s^x(0)\: \mathrm{Im}\left[ \partial_x z(x = 0)\right] 
\propto l_s/l_R$. That is, even if the polarization of the spins at $x = 0$ is along the $x$-axis and therefore
$s^z(0)  = 0$, the gradient of $s^z(x)$ at $x = 0$  does not vanish and thus the contribution of the 
SHE is nonzero. This is also visible (although less clearly) in 
panels Fig.~\ref{fig:Rnl-x decomposition}(d) and (e).

A few other interesting features of  Fig.~\ref{fig:Rnl-x decomposition}  are noteworthy. 
For $\vec{\hat{n}}_{p}=\vec{\hat{z}}$ (left panels),
as the spin Hall angle is increased from  $\theta_{\mathrm{sH}}=0.01$ (panel a)  to $\theta_{\mathrm{sH}}=0.1$ (panel b)
while keeping $l_s/l_{\mathrm{DMC}}$ constant, the non-monotonic behavior of  $\Delta R_{nl}(x)$ disappears.
Indeed, even though the SHE dominates at  distances  $x \lesssim l_s$ for small spin Hall angle, 
as noted above, the contribution arising from  CISP, which is  small for $x \lesssim l_s$  
becomes comparable to the SHE contribution for $x \approx l_s$. 
This is because spins  at $x\sim l_s$ spins  have undergone relaxation and 
precession under the Rashba field onto the plane
where the DMC mechanism is most effective. However, 
as the spin Hall angle is increased to $\theta_{\mathrm{sH}} = -0.1$ (panel b), 
the contribution from the SHE becomes an order of magnitude larger and it is dominant even for $x \sim l_s$.
Thus, the peak in $\Delta R_{nl}(x)$, which results from CISP taking over SHE for $x\sim l_s$, disappears. 
Finally, at the bottom panel (c) of Fig.~\ref{fig:Rnl-x decomposition}, we show results with a decreased 
ratio $l_R/l_s = 2$, which implies that for $x/l_s \sim 1$  the spins undergo a sizable precession in  the 
Rashba field. This enhances the EE contribution to the CISP, which now shows a quantitatively different
behavior from panels (a) and (b). For the plots on the 
right, the spins are injected in plane (along the $x$-axis,
and CISP essentially accounts for most of
the nonlocal  resistance of the device, even though 
for the bottom panel ($l_R/l_s = 2)$  the Rashba precession
gives rise to a sizable contribution from the SHE for $x \lesssim l_s$.

 Finally, let us briefly discuss the effect of the applied magnetic field. The
dimensionless function $f(\vec{\hat{n}}_{p},\omega_{L})$ takes the following forms when $\vec{n}_{p}$
points along the $x$ and $z$ directions, respectively:

\begin{align}
f(\vec{\hat{z}},\omega_{L})=\left[-\theta_{\textrm{sH}}\tilde{q}l_{s}-\frac{l_{s}}{l_{\textrm{DMC}}}\sin\theta_{L}\right]\cos\left(\frac{x}{l_{\textrm{eff}}}\right)\nonumber \\
+\frac{l_{s}}{l_{\textrm{DMC}}}\cos\theta_{L}\sin\left(\frac{x}{l_{\textrm{eff}}}\right)\\
f(\vec{\hat{x}},\omega_{L})=\left[-\theta_{\textrm{sH}}\tilde{q}l_{s}-\frac{l_{s}}{l_{\textrm{DMC}}}\sin\theta_{L}\right]\sin\left(\frac{x}{l_{\textrm{eff}}}\right)\nonumber \\
-\frac{l_{s}}{l_{\textrm{DMC}}}\cos\theta_{L}\cos\left(\frac{x}{l_{\textrm{eff}}}\right)
\end{align}
where  $l_{\textrm{eff}}^{-1}=l_{R}^{-1}-\tilde{q}\sin\theta_{L}$. Thus,
at short distances, $\Delta R_{nl}\sim \theta_{\textrm{sH}}+\left(\sin\theta_{L}/\tilde{q}l_{s}\right)l_{s}/l_{\textrm{DMC}}$
for $\vec{\hat{n}}_{p}=\vec{\hat{z}}$.  On the other hand, $\Delta R_{nl}\sim\cos\theta_{L}\left(l_{s}/l_{\textrm{DMC}}\right)$ for $\vec{\hat{n}}_{p}=\vec{\hat{x}}$. Recall that $\theta_{L}\approx \tfrac{1}{2} \tan^{-1}\left(\omega_{L}\tau_{s}\right)$,
which means that the dominant mechanism at short distance  is  modified (relative to $\omega_L = 0$) by the Larmor precession in the external magnetic field, as expected.

In Fig.~\ref{fig:Nonlocal resistance versus external in-plane magnetic field}, we plot $\Delta R_{nl}$ versus the magnitude of applied magnetic field measured in units of the Larmor frequency times the spin relaxation time, i.e.  $\omega_{L}\tau_s$. For $\vec{\hat{n}}_{p}=\vec{\hat{z}}$, $\Delta R_{nl}$ is almost symmetric because the SHE  contribution dominates over CISP. On the other hand, $\Delta R_{nl}$ becomes almost anti-symmetric asymmetric when the CISP  contribution dominates over the SHE.  For $\vec{\hat{n}}_{p}=\vec{\hat{x}}$, $\Delta R_{nl}$ is highly symmetric when CISP  dominates over SHE (i.e. for $\theta_{\mathrm{sH}}\gg l_s/l_{\mathrm{DMC}}$, while $\Delta R_{nl}$ is highly asymmetric in the opposite limit  where SHE  dominates over CISP.  Thus, in summary, the symmetry of this curve, combined with the very different behavior of  $\Delta R_{nl}(x)$ as a function of the distance $x$ to the injection contact  for zero magnetic field  
and different polarization of the injected spins should provide a ``smoking gun'' for the 
dominant spin-charge conversion mechanism in lateral  spin-valve devices.  

\section{Purely extrinsic SOC} \label{sec:boltzmann} 

In this section, we derive the set of drift-diffusion equations introduced in Sec.~\ref{sec:DDE} from
a model that assumes purely extrinsic SOC. This model is appropriate to graphene decorated
with absorbates. We treat  scattering with the absorbates nonperturbatively, which allows
to describe resonant scattering effects. The latter is very important in graphene due to appearance
of scattering resonances  in the neighborhood of the Dirac point. 
This approximation is valid in the limit of a dilute number of scatterers.
 
We shall rely on the (linearized) quantum Boltzmann equation (QBE) that describes the dynamics of the 2-by-2
density matrix distribution $n_{k}\left(\vec{r},t\right)$ in spin space and reads:
\begin{align}
\partial_{t}\delta n_{k}\left(\vec{r},t\right)+\left(\boldsymbol{v}_{k}\cdot\partial_{\vec{r}}\right)\delta n_{k}\left(\vec{r},t\right)\nonumber \\
+\frac{i}{\hbar}\gamma\left[\delta n_{k}\left(\vec{r},t\right),\boldsymbol{s}\cdot\vec{\mathcal{H}}\left(t\right)\right]+\nonumber \\
e\boldsymbol{E}\left(t\right)\cdot\frac{\nabla_{k}n_{k}^{0}}{\hbar}=\mathcal{I}\left[\delta n_{k}\right],\label{eq:QBE}
\end{align}
In the above expression, the spin operator is given by $\boldsymbol{s}=\frac{\hbar}{2}\boldsymbol{\sigma}$
where $\boldsymbol{\sigma}$ is the Pauli matrices, and the deviation
of the distribution from equilibrium is given by $\delta n_{k}\left(\vec{r},t\right)=n_{k}\left(\vec{r},t\right)-n_{k}^{0}$,
where $n_{k}^{0}=f_{\textrm{FD}}\left[\varepsilon_{k}\right]\mathbb{1}$,
$f_{\textrm{FD}}(\epsilon) = \left[e^{(\epsilon-\bar{\mu})/T}+1\right]^{-1}$ the Fermi-Dirac
distribution at temperature $T$ and chemical potential $\bar{\mu}$,
and $a\mathbb{1}$ is the $2\times2$ identity matrix in spin space. For graphene, 
the dispersion relation for electron is given by $\varepsilon_{k}=\hbar v_{F}k$,
$\boldsymbol{E}\left(t\right)$ is the applied electric field, and
$\vec{\mathcal{H}}\left(t\right)$ is the applied magnetic
field.

The collision integral in the above QBE was derived in Ref.~\citep{Huang_PRB2016}
to leading order in the density of impurities, and reads:$n_{\textrm{imp}}$, is given by the following expression: 
\begin{align}
\mathcal{I}\left[\delta n_{k}\right]=\frac{i}{\hbar}\left[\delta n_{k},\mathrm{Re}\, \Sigma_{k}\right]+\frac{2\pi n_{\textrm{imp}}}{\hbar}\sum_{p}\delta\left(\epsilon_{k}-\epsilon_{p}\right)\nonumber \\
\times\left[T_{kp}^{+}\delta n_{p}T_{pk}^{-}-\frac{1}{2}\left\{ \delta n_{k},T_{kp}^{+}T_{pk}^{-}\right\} \right]\label{coll-int}
\end{align}
The self energy $\Sigma_{k}^{R}$  reads as
\begin{equation}
\mathrm{Re}\,\Sigma_{k}^{R}=\frac{n_{\textrm{imp}}}{2}\left(T_{kk}^{+}+T_{kk}^{-}\right)
\end{equation}
In order to derive the drift-diffusion equations, we use the following ansatz to
solve the QBE:

\begin{multline}
n_{k}^{0}+\delta n_{k}\left(\vec{r},t\right)=f_{\textrm{FD}}\left[\varepsilon_{k}-\mu\left(\vec{r},t\right)-h_{0}\boldsymbol{\sigma}\cdot\boldsymbol{n}_{0}\left(\vec{r},t\right)\right. \\
\left.-\hbar\boldsymbol{k}\cdot\boldsymbol{v}_{c}\left(\vec{r},t\right)-\hbar\boldsymbol{k}\cdot\boldsymbol{v}_{s}\left(\vec{r},t\right)\left(\boldsymbol{\sigma}\cdot\boldsymbol{n}_{1}\left(\vec{r},t\right)\right)\right]\label{eq:n-ansatz}
\end{multline}
In what follows, we shall look for a solution of the QBE to linear 
order in $\mu, h_0, \vec{v}_c,\vec{v}_s$,
and $\mu$. Here $\mu(\vec{r},t)$  is the local deviation from the average 
chemical potential, $\bar{\mu}$; 
$\boldsymbol{v}_{c}(\vec{r},t)$ ($\boldsymbol{v}_{\boldsymbol{s}}(\vec{r},t)$)
is the local drift velocity of the charge (spin); $\boldsymbol{n}_{0}(\vec{r},t)$ ($\boldsymbol{n}_{1}(\vec{r},t)$)
is the polarization direction of the nonequilibrium magnetization  (spin current). 
The parameters in the above ansatz are 
related to the charge density $\rho\left(\vec{r},t\right)$,
spin density $\boldsymbol{s}\left(\vec{r},t\right)$, charge density
current $\boldsymbol{J}\left(\vec{r},t\right)$, and spin current density
$\boldsymbol{\mathcal{J}}^{a}\left(\vec{r},t\right)$ by the following expressions:
\begin{align}
\rho\left(\vec{r},t\right)&=\frac{g_{s}g_{v}}{2\Omega}\sum_{k}\mathrm{Tr}\left[\delta n_{k}\left(\vec{r},t\right)\right]\notag\\
&=g_{s}g_{v}N_{0}\mu\left(\vec{r},t\right)\label{eq:charge}
\end{align}
\begin{align}
\vec{s}\left(\vec{r},t\right)&=\frac{g_{s}g_{v}}{2\Omega}\sum_{k}\mathrm{Tr}\left[\vec{\sigma}\delta n_{k}\left(\vec{r},t\right)\right]
\notag\\
&=g_{s}g_{v}N_{0}h_{0}\vec{n}_{0}\left(\vec{r},t\right)
\end{align}
\begin{align}
\boldsymbol{J}\left(\vec{r},t\right)&=\frac{g_{s}g_{v}}{2\Omega}\sum_{k}\mathrm{Tr}\left[\delta n_{k}\left(\vec{r},t\right)\right]\boldsymbol{v}_{k}\notag\\
&=g_{s}g_{v}\frac{N_{0}}{2}\varepsilon_{F}\boldsymbol{v}_{c}\left(\vec{r},t\right)
\label{eq:charge-current}
\end{align}
\begin{align}
\boldsymbol{\mathcal{J}}^{a}\left(\vec{r},t\right)&=\frac{g_{s}g_{v}}{2\Omega}\sum_{k}\mathrm{Tr}\left[\sigma^{a}\delta n_{k}\left(\vec{r},t\right)\right]\boldsymbol{v}_{k}\nonumber \\
&=g_{s}g_{v}\frac{N_{0}}{2}\varepsilon_{F}\boldsymbol{v}_{s}\left(\vec{r},t\right)n_{1}^{a}\left(\vec{r},t\right)
\label{eq:spin-current-def}
\end{align}
Here $g_{s}$ and $g_{v}$ are spin degeneracies and valley degeneracies
receptively, $N_{0}$ is the density of states per spin per valley
at the Fermi surface. In evaluating the sums over momentum above, we have assumed the
low-temperature limit where $T\ll \bar{\mu}$ and approximated $\partial_{\epsilon}n_{k}^{0}\simeq
-\delta(\epsilon_{k}-\epsilon_F)$ where $\varepsilon_F = \bar{\mu}(T = 0)$ is the Fermi energy.

Note in Eq.~\eqref{eq:charge-current} and \eqref{eq:spin-current-def},
the currents are given by the first moment of deviation from 
equilibrium of the distribtion function. In the presence of SOC, they are \textit{not} the conserved
current that enters the continuity equation. The conserved current
is a sum of two distinct contributions: the first moment excitation
of the Fermi surface and the anomalous current which arised from evaluating
the collision integral to order $k_{F}^{-1}\nabla_{r}$ \citep{Bergeret_PRB2016}.
In fact, the anomalous current contributes precisely to the so-called
side-jump contribution, see Ref.~\citep{Huang_PRB2018} for more
in-depth discussion. However, if we limit ourselves to study spin-charge
coefficients to the leading order in impurity density $n_{\textrm{imp}}$,
the collision integral in Eq.~\eqref{coll-int} is sufficient and
the conserved currents are still given by Eq.~\eqref{eq:charge-current}
and \eqref{eq:spin-current-def}.

Next, we comopute the (retarded) $T$-matrix for a single impurity. The latter
is a $2\times2$ matrix in spin space, which can written as follows:
\begin{equation} \label{eq:T-retarded}
T_{kp}^{+}=C_{kp}\mathbb{1}+\boldsymbol{B}_{kp}\cdot\boldsymbol{\sigma}
\end{equation}
where the coefficients $C_{kp}$ and $\boldsymbol{B}_{kp}$ are given by: 
\begin{align}
C_{kp}=\gamma_{0}\cos \left( \frac{\theta_{k}-\theta_{p}}{2} \right)
\end{align}
\begin{align}
\boldsymbol{B}_{kp}=&\gamma_{R}\sin\left( \frac{\theta_{k}+\theta_{p}}{2} \right)\hat{x}-\gamma_{R}\cos \left( \frac{\theta_{k}+\theta_{p}}{2} \right) \hat{y} \notag \\
&+i\gamma_{I}\sin \left( \frac{\theta_{k}-\theta_{p}}{2} \right) \hat{z}
\end{align}
 This parametrization
of $T$-matrix follows from symmetry considerations. It respects the rotation
generated by total angular momentum (spin angular momentum + orbital
angular momentum), in-plane parity and time-reversal symmetry but
breaks $z\rightarrow-z$ symmetry.

For a given single-impurity $T$-matrix, the equations
of motion for  the different moments of the distribution function 
(Eq.~\eqref{eq:charge}-\eqref{eq:spin-current-def}) can be obtained to leading
order in the impurity density.  This involves taking the zeroth and first moments  
of Eq.~\eqref{eq:QBE} followed by the the trace of the result over the
spin indices. Those manipulations yield the following set of equations:
\begin{widetext}
\begin{align}
\partial_{t}\rho\left(\vec{r},t\right)+\partial_{i} J_{i}\left(\vec{r},t\right)=0
\end{align}
\begin{align}
\partial_{t}\boldsymbol{s}\left(\vec{r},t\right)+\partial_{i}\boldsymbol{\mathcal{J}}_{i}\left(\vec{r},t\right)+\gamma\boldsymbol{\mathcal{H}}\left(t\right)\times\boldsymbol{s}\left(\vec{r},t\right)=\boldsymbol{\mathcal{Q}}(\vec{r},t)\label{eq:spin_density evo}
\end{align}
\begin{align}
\partial_{t}J_{i}\left(\vec{r},t\right)+\frac{v_{F}^{2}}{2}\partial_{i}\rho\left(\vec{r},t\right)-\frac{\sigma_{D}}{\tau_{c}}E_{i}\left(t\right)=-\frac{J_{i}\left(\vec{r},t\right)}{\tau_{c}}+\alpha_{\textrm{sk}}\varepsilon_{ij}\mathcal{J}_{j}^{z}\left(\vec{r},t\right)+\alpha_{\textrm{asp}}v_{F}\varepsilon_{ij}s^{j}\left(\vec{r},t\right)\label{eq:charge-current_density evo}
\end{align}
\begin{align}
\partial_{t}\mathcal{J}_{i}^{a}\left(\vec{r},t\right)+\frac{v_{F}^{2}}{2}\partial_{i}s^{a}\left(\vec{r},t\right)+\gamma\left[\vec{\mathcal{H}}\left(t\right)\times\boldsymbol{\mathcal{J}}_{i}\left(\vec{r},t\right)\right]^{a}=\chi_{i}^{a}(\vec{r},t)\label{eq:spin_current_density evo}
\end{align}
\end{widetext}
The components of $\boldsymbol{\mathcal{Q}}(\vec{r},t)$ and $\chi_{i}^{a}(\vec{r},t)$,
as well as the scattering rates are given in Appendix~\ref{app:scattering rates}.

To proceed further, we set $\partial_{t}J_{i}=\partial_{t}\mathcal{J}_{i}^{a}=0$ as corresponds
to the steady state. Hence, the constitutive relations for the charge $J_{i}\left(\vec{r}\right)$
and spin  $\mathcal{J}_{i}^{a}\left(\vec{r}\right)$  current densities are
derived from the Eqs.~\eqref{eq:charge-current_density evo}
and~\eqref{eq:spin_current_density evo}:

\begin{align}
J_{i}=-D\partial_{i}\rho+\sigma_{D}E_{i}+\theta_{\textrm{sH}}\varepsilon_{ij}\mathcal{J}_{j}^{z}+\alpha_{\textrm{asp}}\tau_{c}v_{F}\varepsilon_{ij}s^{j}
\end{align}
\begin{align}
\mathcal{J}_{i}^{z}=-D\partial_{i}s^{z}+\theta_{\textrm{sH}}\varepsilon_{ij}J_{j}+\alpha_{\textrm{R}}\tau_{c}v_{F}s^{i}
\end{align}
\begin{align}
\mathcal{J}_{x}^{x}=-D^{\prime}\partial_{x}s^{x}-\alpha_{\textrm{R}}^{\perp}\tau_{c}^{\prime}v_{F}s^{z}-\alpha_{\textrm{LD}}\tau_{c}^{\prime}\mathcal{J}_{y}^{y}
\end{align}
\begin{align}
\mathcal{J}_{y}^{y}=-D^{\prime}\partial_{y}s^{y}-\alpha_{\textrm{R}}^{\perp}\tau_{c}^{\prime}v_{F}s^{z}-\alpha_{\textrm{LD}}\tau_{c}^{\prime}\mathcal{J}_{x}^{x}
\end{align}
\begin{align}
\mathcal{J}_{x}^{y}=-D^{\prime\prime}\partial_{x}s^{y}+\alpha_{\textrm{LD}}^{\perp}\tau_{c}^{\prime\prime}\mathcal{J}_{y}^{x}
\end{align}
\begin{align}
\mathcal{J}_{y}^{x}=-D^{\prime\prime}\partial_{y}s^{x}+\alpha_{\textrm{LD}}^{\perp}\tau_{c}^{\prime\prime}\mathcal{J}_{x}^{y}
\end{align}
Here $\theta_{\textrm{sH}}=\alpha_{\textrm{sk}}\tau_{c}$ is the
spin-Hall angle, and the diffusion constants are given by $D=\frac{1}{2}v_{F}^{2}\tau_{c}$,
$D^{\prime}=\frac{1}{2}v_{F}^{2}\tau_{c}^{\prime}$, $D^{\prime\prime}=\frac{1}{2}v_{F}^{2}\tau_{c}^{\prime\prime}$.

In order to further simplify the calculations, we shall  take  $\tau_{c}=\tau_{c}^{\prime}=\tau_{c}^{\prime\prime}$.
and  $\alpha_{\textrm{R}}=\alpha_{\textrm{R}}^{\perp}$ since they differ by terms that are proportional to the 
SOC induced by the impurities, which are  typically small compared to the scalar potential term. In addition, we shall  drop the terms  proportional to $\alpha_{\textrm{LD}}$ and $\alpha_{\textrm{LD}}^{\perp}$, which describe the
Lifshitz-Dyakonov spin swapping effect~\cite{Lifshits_PRL2009}. For $\alpha_{LD}\tau_c \ll 1$, this effect 
leads to corrections that are second order in the spin-charge conversion coefficients. The latter, as
pointed out above, are typically  smaller than one in spintronic devices. Thus,  second order effects 
are negligible and can be  neglected.  The resulting equations can be brought to the form of 
Eqs.~\eqref{eq:c-consti} and~\eqref{eq:s-consti} with the following choice of parameters: 
\begin{align}
\gamma_{ij}^{a}&=\alpha_{\textrm{sk}}\tau_{c}\epsilon_{ij}\delta^{az}\label{eq:gamma_ij^a QBE}
\\
\\A_{i}^{b}&=\frac{2\alpha_{R}}{v_{F}}\varepsilon^{b}_{\,\,i}=l_{R}^{-1}\varepsilon^{b}_{\,\,i}\label{eq:A_i^b QBE}
\\
\kappa_{i}^{a}&=\frac{2\alpha_{\textrm{asp}}}{v_{F}}\varepsilon_{i}^{\,\,a}=l_{\textrm{DMC}}^{-1}\varepsilon_{i}^{\,\,a}\label{eq:kappa QBE}\\
\Gamma_{s}^{xx,yy}&=\frac{1}{\tau_{\textrm{EY}}} \label{eq:tau_s QBE}\\
\Gamma_{s}^{zz}&=\frac{1}{\tau_{\textrm{EY}}^{\perp}} \label{eq:tau_s^z QBE}
\end{align}
and $\Gamma^{ab} = 0$ for $a\neq b$.
 The detailed forms  of $\alpha_{\textrm{sk}}$, $\alpha_{R}$, $\alpha_{\textrm{asp}}$,
$\tau_{\textrm{EY}}$, $\tau_{\textrm{EY}}^{\perp}$ in terms of the scattering rates
with the impurities are given in Appendix~\ref{app:scattering rates}.

By relying on the one-dimensional approximation introduced in Sec.~\ref{sec:spin-valve},
the diffusion equation for the spin density $\boldsymbol{s}$ in the presence of a 
weak external magnetic field ($\omega_{L}\tau_{c}\ll1$) can be written as follows:
\begin{align}
\bar{\boldsymbol{\mathcal{D}}}\boldsymbol{s}\left(x\right)-\omega_{L}
\left[\vec{\hat{n}}_H\times\boldsymbol{s}\left(x\right)\right]=\boldsymbol{\mathcal{S}}\left(x\right),
\end{align}
where $\boldsymbol{\mathcal{S}}$ is the source term:
\begin{align}
\boldsymbol{\mathcal{S}}\left(x\right)=\left(2\alpha_{\textrm{asp}}\frac{J_{y}\left(x\right)}{v_{F}},-2\alpha_{\textrm{asp}}\frac{J_{x}\left(x\right)}{v_{F}},\theta_{\textrm{sH}}\partial_{x}J_{y}\left(x\right)\right),
\end{align}
The diffusion matrix $\bar{\boldsymbol{\mathcal{D}}}$ is  
\begin{align}
\bar{\boldsymbol{\mathcal{D}}}=\left(\begin{array}{ccc}
D^{\prime}\partial_{x}^{2}-\frac{1}{\tau_{\textrm{EY}}} & 0 & \theta_{\textrm{R}}v_{F}\partial_{x}\\
0 & D^{\prime\prime}\partial_{x}^{2}-\frac{1}{\tau_{\textrm{EY}}} & 0\\
-\theta_{\textrm{R}}v_{F}\partial_{x} & 0 & D\partial_{x}^{2}-\frac{1}{\tau_{\textrm{EY}}^{\perp}}
\end{array}\right),
\end{align}
where $\theta_{R}=\tau_{c}\alpha_{\textrm{R}}+\tau_{c}^{\prime}\alpha_{\textrm{R}}^{\perp}$. The above diffusion matrix can be reduced to Eq.~\eqref{eq: Diffusion matrix isotropic} if we assume $\tau_{\textrm{EY}} = \tau^{\perp}_{\textrm{EY}}$ in order to simplify the model, as explained in Sec.~\ref{sec:spin-valve}. 

 Furthermore, concerning the source term, screening ensures that the charge density is  uniform for length scales larger than the Thomas-Fermi screening length.  Therefore, to leading order in the spin-charge conversion coefficients, the charge current density $J \approx -D\nabla \rho +\sigma_{D}E = 0$  and hence 
$\boldsymbol{\mathcal{S}}(x) = 0$  in the bulk of the device described in Sec.~\ref{sec:spin-valve}.

\section{Intrinsic SOC with random fluctuations}
\label{sec:su2}
In this section, we shall describe the proximity induced SOC as field consisting of a spatially 
uniform (i.e. a `intrinsic' SOC) part and a random component  that varies slowly in space. Thus, 
the  spin-charge diffusion equations can be  derived from a  kinetic theory that treats the  SOC as a non-abelian gauge field~\cite{Shen_PRB2014,Shen_PRL2014,Huang_Milletari_Cazalilla}. Remarkably, the resulting diffusion equations take the universal form as those  introduced in Eq.~\eqref{eq:c-cont} -\eqref{eq:s-consti}. In what follows, we first generalized the celebrated 2D Rashba model  \footnote{In this section, we consider the Rashba model and not the Dirac-Rashba model for graphene since the sub-lattice pseudospin degree of freedom is not important when the Fermi energy is large.} to account for smoothly varying SOC potential then, a gauge-covariant kinetic theory is introduced to derive the diffusion equations.

 Let us consider a 2D electron gas with Rashba SOC (the so-called Rashba model)  and re-write the SOC as time-indenpedent uniform non-abelian gauge-field:
\begin{align} \label{eq:rashba}
H_R=\frac{\vec{p}^2}{2m}+\alpha( \vec{\sigma} \wedge \vec{p})=\sum_{i=x,y}\frac{(p_i- \mathcal{A}_i)^2}{2m} + \text{const}.
\end{align}
Here $\vec{a}\wedge\vec{b} = \epsilon_{ij} a_i b_j$, and  $\alpha$ is the  strength of uniform (intrinsic) part of the SOC whilst $\mathcal{A}_i$ is the non-abelian gauge field:
\begin{equation}
\mathcal{A}_i=\sum_{a=x,y,z}\mathcal{A}_i^a \sigma_a.
\end{equation}
For Rashba SOC, the only non-vanishing components are are $\mathcal{A}_y^x=-\mathcal{A}_x^y=m\alpha$. In the literature on proximity effects in 2D metals, it is often assumed that proximity-induced SOC is uniform in space and therefore $[p_j,\mathcal{A}_i]=0$. Thus,  the violation of momentum conservation that is needed  in order for the system to reach the steady state is assumed to be driven by scattering with  impurities. However, as  emphasized  above, a  realistic SOC induced by proximity should contain both uniform and spatially random components. Thus, in order to account for the random spatial fluctuations,  we have generalized the Rashba model introduced above in Eq.~\eqref{eq:rashba} by introducing an electrostatic potential  $\phi(\vec{r})$ and shifting  the gauge field as $ \mathcal{A}_i\rightarrow\ \mathcal{A}_i +\delta \mathcal{A}_i(\vec{r})$, which yields the following model:
\begin{equation} \label{eq:Hnew}
H=\sum_{i=x,y}\frac{(p_i- \mathcal{A}_i-\delta\mathcal{A}_i(\vec{r}))^2}{2m} + \phi(\vec{r})
\end{equation}
The   potential $\phi(\vec{r})$ is a slowly varying function in space and its spatial variation gives rise to finite electric field that generates SOC. In fact, the spatially varying gauge-field is induced by the gradient of the electrostatic 
potential $\phi(\vec{r})$:
\begin{align}
\delta \mathcal{A}_i^z(\vec{r}) &= m \alpha_1  \epsilon_{ij} \partial_j \phi(\vec{r}),\\
\delta \mathcal{A}_i^j(\vec{r}) &= m \alpha_2 \epsilon_{i}^{\,\,j} \partial_z \phi(\vec{r})
\end{align}
Here $\partial_z \phi = \partial_z \phi(\vec{r},z)|_{z=0}$ where $z=0$ is the material plane; $\alpha_1 \sim \alpha$ ($\alpha_2 \sim \alpha$) are material-dependent coefficients that characterize the strength of SOC induced by in-plane (out-of-plane) electric field ($\vec{E}=-\nabla \phi$). Note that the generalized Hamiltonian Eq.~\eqref{eq:Hnew} breaks translational symmetry but retains all other symmetries of the Rashba Hamiltonian Eq.~\eqref{eq:rashba}. 

  In order to proceed further, it is convenient to isolate the part that breaks translation symmetry from the Rashba Hamiltonian: $H=H_R+U(\vec{r},\vec{p})$ where $H_R$ is given in Eq.~\eqref{eq:rashba} and
\begin{equation}  \label{eq:U(rp)}
U(\vec{r},\vec{p})= - \frac{1}{2m}\{  p_i \, ,\, \delta \mathcal{A}_i(\vec{r})\} +
\phi(\vec{r}).
\end{equation}
We have dropped the subleading  term $\propto (\delta\mathcal{A}_i)^2$  since it is  $\sim \alpha^2$ and small compared to the other two. The  matrix elements of this potential are:
\begin{equation} \label{eq:U_dis}
U_{ \vec{k} \vec{p}}= \phi_{\vec{k}- \vec{p}} \left\{ 1+ i  \alpha_{1} \left(\vec{k} \wedge \vec{p}\right)\sigma^z - \frac{\alpha_{2}}{2\xi} \left[ \left(\vec{p}+\vec{k}\right) \wedge  \boldsymbol{\sigma} \right] \right\},
\end{equation}
where $\phi_{\vec{k}-\vec{p}}$ is the Fourier component of the electric potential and we have approximated $\partial_z \phi\approx \phi/ \xi$. Here $\xi$ is a typical length scale of variation in the direction out of the 2D plane. The resulting potential is similar to those described in Refs.~\cite{sherman-1,sherman-2,Huang_Milletari_Cazalilla},

We shall consider the situation where both the fluctuating and uniform components of the SOC are small compared to the Fermi energy $\alpha_1 p_F^2  \sim \alpha_2 p_F/\xi \sim \alpha /v_F  \ll 1 $. In this limit, starting from the structure of Eq.~\eqref{eq:Hnew}, one can write down a kinetic equation for the (spin) density-matrix distribution function $n_{\vec{k}}(\vec{r},t)$ by relying on gauge invariance (cf.  Ref.~\cite{Raimondi_Annalen2012,Shen_PRB2014,Bergeret_PRB2014}):
\begin{align}
\left( \nabla_{t} \,n_{\vec{k}} +  \boldsymbol{v}_{\vec{k}} \cdot   \nabla_{\vec{r}} n_{\vec{k}}  \right) +\frac{1}{2}\left\{\vec{F}_{\vec{k}} , \partial_{\vec{k}} n_{\vec{k}} \right\} 
= I[\delta n_{\vec{k}}]. \label{eq:QBE_general}
\end{align}
The intrinsic SOC (i.e. the non-abelian gauge field) modifies the left hand side (dissipation-less part) of the kinetic equation in two essential ways:
First,  it turns the space-time derivatives into covariant derivatives: $\nabla_{\vec{r}}$ ($\nabla_{t}$) is the covariant space (time) derivative that describes the precession of electron spin induced by SOC  (external magnetic field).  Mathematically, 
the covariant derivatives on the right hand-side of the kinetic equation have a structure is identical to Eq.~\ref{eq:cov-der}.
However, as we shall see later, the  non-abelian gauge connections are renormalized by the fluctuating part of the SOC.
Second, $\vec{F}_{\vec{k}}$ is the non-abelian generalization of external applied force acting on electron $\vec{k}$. The three spatial components of the non-abelian force are obtained from $F^j_{\vec{k}}=\text{V}_{a} \mathcal{F}^{a j}$ where $(\text{V}_{a})=(1, v_{x\vec{k}},v_{y\vec{k}},0)$ is the four-velocity and $\mathcal{F}^{a b}=\partial^{a} \mathcal{A}^b - \partial^{b} \mathcal{A}^{a}-[\mathcal{A}^a, \mathcal{A}^b]$ is the field strength tensor. Here the indices $j=x,y,z$ while the indices $a,b=t,x,y,z$.  For example, if we submit an electric field $\vec{E}$ in the presence of Rashba SOC with gauge-field $\mathcal{A}^x_y=-\mathcal{A}^x_y=m\alpha$, the resulting non-abelian force contains a spin-dependent Lorentz force responsible for the  intrinsic spin Hall effect \footnote{Note this intrinsic spin Hall effect is \textit{not} a result of summation of the band Berry curvature.}: 
\begin{equation}
\vec{F}_{\vec{k}} = e \, \vec{E}+ \vec{v}_{\vec{k}} \times \left(e \, \boldsymbol{\mathcal{B}}_s\right)
\end{equation}
where $\boldsymbol{\mathcal{B}}_s = (8 m^2 \alpha^2/e^2) \sigma^z \vec{\hat{z}} $ is the spin-dependent magnetic.

The potential  $\phi(\vec{r})$ is treated as a random potential, which contributes to the relaxation of
momentum and spin and therefore must described by the collision integral of the kinetic equation.  
The collision integral to second order in $\delta \mathcal{A}$, in the self-consistent Born-approximation, 
takes the form:
\begin{align}
\mathcal{I}\left[\delta n_{k}\right]=\frac{i}{\hbar}\left[\delta n_{k},\mathrm{Re}\, \Sigma_{k}^{\mathrm{B}}\right]+\frac{2\pi }{\hbar}\sum_{p}\delta\left(\epsilon_{k}-\epsilon_{p}\right)\nonumber \\
\times\left[ \overline{U_{kp}\delta n_{p}U_{pk}} - \frac{1}{2}\left\{ \delta n_{k}, \overline{ U_{kp}U_{pk} }\right\} \right],
\label{coll-int}
\end{align}
where $\Sigma_{k}^{\mathrm{B}}$ is the hermitian part of the self-energy:
\begin{equation}
\mathrm{Re}\, \Sigma_{k}^{\mathrm{B}}= \overline{U}_{kk}+ P\int \frac{d^2 q}{(2\pi)^2} \frac{\overline{U_{kq}U_{qk}}}{\epsilon-\epsilon_q} \label{eq:born-sigma}
\end{equation}
Here $\overline{ O[\phi]}= \sum_{\phi} P[\phi] O[\phi]$ and  $P[\phi]$ is the probability distribution function of the random potential $\phi$. For simplicity, we assume they are distributed according to Gaussian distribution with zero mean:
\begin{align}
\overline{ \phi_{\vec{q}} }=& 0 \\
\overline{ \phi_{\vec{q}_1}\phi_{\vec{q}_2} }=& n_s v_0^2\, \delta^{2}(\vec{q_}1 + \vec{q}_2)  
\end{align}
The parameter $n_s$ has dimensions of inverse length square and is akin to  $n_{imp}$ in Sec.~\ref{sec:boltzmann};
 $v_0$ is the typical  energy scale of the random part of the proximity induced electric potential $\phi(\vec{r})$. Since $\phi(\vec{r})$ has zero mean value, the first term in Eq.~\eqref{eq:born-sigma} vanishes under potential average. However the second term does not vanish and still contributes to the energy shift. 
Then, unlike the uniform gauge field $\mathcal{A}_i$, the fluctuating gauge-field $\delta \mathcal{A}_i$ generates dissipation and enters the kinetic theory via the collision integral. 
For a potential $\phi(\vec{r})$ with short-range correlations, the collision integral in Eq.~\eqref{coll-int} suffices to describe  the spin-charge relaxation since it accounts for the matrix structure of the disorder potential, i.e.~Eq.~\eqref{eq:U(rp)}. However, it is still an approximation because Eq.~\eqref{coll-int} does not account for the  modification of the scattering states by the uniform  part of the SOC $\mathcal{A}_i\sim \alpha$: The asymptotic scattering states are given by spin-independent Bloch waves with energy $\epsilon_k=v_F k$.  This is consistent with our assumption of a weak SOC with our treatment of the left-hand side of Eq.~\eqref{eq:QBE_general}, which is valid to second order in $\alpha$.

 After using the same ansatz as in Eq.~\eqref{eq:n-ansatz} to solve the above kinetic equation, we arrive at the set of drift-diffusion equations, Eqs.~\eqref{eq:c-cont} to \eqref{eq:s-consti} with the following identification for the parameters:
\begin{equation}
\gamma_{ij}^{a}=\frac{8m\alpha^{2}}{\pi n_{s}N_{0}v_{0}^{2}}
 \frac{ \epsilon_{ij}\delta^{az} }
{\left(2+\alpha_{1}^{2}k_{F}^{4}+2\left(\frac{\alpha_{2}}{2\xi}\right)^{2}k_{F}^{2}\right)}
\end{equation}
\begin{equation}
A_{i}^{b}=\left[\mathcal{A}_{y}^{x}-\frac{4mn_{s}}{\pi\hbar v_{F}}v_{0}^{2}\left(\frac{\alpha_{2}}{2\xi}\right)^{2}\ln\left(\frac{q_{c}}{k_{F}}\right)\right]\epsilon^{b}_{\,\,i}
\end{equation}
\begin{equation}
\kappa_{i}^{a}=\frac{4\pi n_{s}}{\hbar v_{F}}N_{0}v_{0}^{2}\alpha_{1}\left(\frac{\alpha_{2}}{2\xi}\right)k_{F}^{3}\epsilon_{i}^{\,\,a}
\end{equation}
\begin{align}
\Gamma^{xx,yy}_s &= \frac{1}{\tau_{s}^{x,y}}\\
&=\frac{2\pi n_{s}}{\hbar}N_{0}v_{0}^{2}\left[2\left(\frac{\alpha_{2}}{2\xi}\right)^{2}k_{F}^{2}+\alpha_{1}^{2}k_{F}^{4}\right]
\end{align}
\begin{equation}
\Gamma^{zz}_{s} = \frac{1}{\tau_{s}^{z}}=\frac{8\pi n_{s}}{\hbar}N_{0}v_{0}^{2}\left(\frac{\alpha_{2}}{2\xi}\right)^{2}k_{F}^{2}
\end{equation}
In the above equations, $k_F$ is the Fermi momentum, and $q_c\sim k_F$ is  high-momentum cut-off. Note that the total gauge-field $A_i^b$ appearing in the diffusion equation receives contributions from both the uniform gauge field ($\mathcal{A}_{y}^{x}$) and the fluctuating gauge field ($\delta A\propto n_s v_0^2$).

\section{Summary}

In this work, we have extended the theory of spin-injection in 2D metals to account for proximity induced spin-orbit coupling (SOC). The theory relies on a set of  diffusion equations that capture the two main types of mechanisms for spin-charge conversion, namely the spin Hall effect (SHE) and the current-induced spin polarization (CSIP). For the latter, two kinds
of contributions have been identified and accounted for: the Edelstein effect, which generates a spin polarization via
the SHE coupled with spin precession caused by the Rashba SOC, and the direct magneto electric coupling (DMC). 
The latter describes a direct coupling between the spin polarization and the electric current, which can arise in systems
with random SOC. We would like to emphasize that such random SOC should be generically present in 2D metals
with proximity induced SOC.

Our calculations for a lateral spin-valve device allowed us to identify the SHE and CSIP  contributions to the non-local 
resistance of the device. Thus, we have been able to ascertain the conditions under which, by changing the quantization axis of the injected spins, the observed nonlocal signal is dominated by one of the two spin-charge conversion mechanism
mentioned above.

In addition, we have provided a microscopic derivation of the  diffusion equations. This has been achieved by treating the describing the proximity-induced SOC in two physically distinct limits. In one of them, we have assumed that SOC is induced by spatially localized impurities. This limit is applicable e.g. to graphene randomly decorated with absorbates (or clusters thereof). In the other limit, we have assumed that SOC consist of a uniform part plus a random component, which is appropriate to 2D heterostructures of  graphene or another two-dimensional metal placed on transition metal
dichalcogenides, for instance.  We have show that the resulting set of equations is identical, which  suggests that the coupled spin-charge diffusive equations derived here apply to a broad class of 2D materials in the metallic regime.

 The theory presented here can be extended in a number of directions: For instance,  accounting for the anisotropy 
 in the spin relaxation should be relatively easy at the expense of introducing an additional (anisotropy) parameter, and also for a moderate spin valley coupling in graphene/TMD heterostructures  which can be described as a valley dependent
Zeeman coupling.

\acknowledgments
MAC and YHL have been supported by the Ministry of Science and Technology (Taiwan) under contract number NSC 102- 2112-M-007-024-MY5. MAC also acknowledges the support of the National Center for Theoretical Sciences of Taiwan. A.F. gratefully acknowledges the financial support from the Royal Society, London through a Royal Society University Research Fellowship. M.O. and A.F. acknowledge funding from EPSRC (Grant Ref: EP/N004817/1).

\appendix

\section{Solution of the Bloch equation}
\label{app:spin-valve}

In this section, we provide the details of the calculation leading
to the dimensionless parameters, $C_{\mathrm{inj}}$ and $f(\vec{\hat{n}}_{p},\omega_{L})$,
is given. The  solution to the spin-diffusion equation, Eq.~\eqref{eq: Diffusion equation}
is displayed in Eqs.~\eqref{eq:soldiff}. The equation for $s^y$ is decoupled from those 
of $s^x$ and $s^z$ and its solution reads $s^y(x) = s^y(0) e^{-x/l_s}$.

Since the injected spin of polarization is along the polarization direction
$\vec{\hat{n}}_{p}$ of the ferromagnet, the problem of enforcing the
boundary conditions  (cf. Eqs.~\eqref{eq:bc1} and \eqref{eq:bc2})  
is largely simplified by projecting the spin current density along $\vec{\hat{n}}_{p}$
on both sides of the ferromagnet-2D material junction, i.e. 
\begin{align}
\mathcal{J}_{N}\left(x=0\right)&=\sum_{\sigma}\sigma\mathcal{J}_{N}^{\sigma}\left(x=0\right)\notag\\
&\approx-2D\vec{\hat{n}}_{p}\cdot\partial_{x}\boldsymbol{s}\left(x=0\right)
\end{align}
\begin{align}
\mathcal{J}_{F}\left(z=0\right)=\sum_{\sigma}\sigma\mathcal{J}_{F}^{\sigma}\left(z=0\right)
\end{align}
Here $\mathcal{J}_{N}^{\sigma}\left(x=0\right)$ and $\mathcal{J}_{F}^{\sigma}\left(z=0\right)$
are the spin current density in the channel $\sigma = \pm 1$ ($+ \equiv \uparrow, - \equiv \downarrow$), 
which points in the direction $\sigma\vec{\hat{n}}_{p}$. Note that we neglect any
interfacial spin-flip scattering, so that the polarization of the
total spin-current flowing into the 2D metal is parallel to the polarization
of the spin current in the ferromagnet: 
\begin{align}
\left[\boldsymbol{\mathcal{J}}_{N}\left(x=0^{+}\right)-\boldsymbol{\mathcal{J}}_{N}\left(x=0^{-}\right)\right]\parallel\vec{\hat{n}}_{p}\label{eq:inject_parallel}
\end{align}

  Since non-local resistance must depend on several
junction properties such as interfacial conductance, interfacial current
polarization, and the current polarization within the ferromagnetic
metal, we construct the following electrochemical potential model
with two channels pointing in $\pm\vec{\hat{n}}_{p}$ direction respectively
in ferromagnetic metal and 2D metal in order to capture the influence
of junction properties:
\begin{align}
\mu_{N}^{\sigma}\left(x\right)=\bar{\mu}_{N}\left(x\right)+\frac{\sigma}{2N_{e}}\boldsymbol{s}\left(x\right)\cdot\vec{\hat{n}}_{p}\label{eq:chemical_potential N}
\end{align}
\begin{align}
\mu_{F}^{\sigma}\left(z\right)=\frac{e^{2}I}{\sigma_{F}A_{J}}z+eV_{1}+b\sigma\left(\frac{\sigma_{F}}{\sigma_{F}^{\sigma}}\right)e^{-z/\lambda_{F}},\label{eq:chemical_potential F}
\end{align}
where  $\bar{\mu}_{N}\left(x\right)=\frac{e^{2}I}{w\sigma_{N}}x$ for
$x<0$, $\bar{\mu}_{N}\left(x\right)=0$ for $x>0$, $V_{1}$ is the
voltage drop between the ferromagnet and the 2D metal, $A_{J}$ is
the cross section of the ferromagnetic metal, $N_{e}$ is the density
of states per spin when the system is at equilibrium, $\lambda_{F}$
is the spin-diffusion length in the ferromagnet, $\sigma_{F}^{\sigma}$
is the spin-dependent electric conductivity of the ferromagnet,
and $\sigma_{F}=\sigma_{F}^{\uparrow}+\sigma_{F}^{\downarrow}$ is
the total electric conductivity in the ferromagnet. The electrochemical
potential Eqs.~\eqref{eq:chemical_potential N} and~\eqref{eq:chemical_potential F}
are constructed within the guideline that the spin current density
projected onto channel $\sigma$ should be given by the following:
\begin{align}
\boldsymbol{\mathcal{J}}_{N\left(F\right)}^{\sigma}=-\frac{\sigma_{N\left(F\right)}^{\sigma}}{e}\partial_{\vec{r}}\mu_{N\left(F\right)}^{\sigma}
\end{align}

To proceed further, we assume that the spin current projected onto
the quantum axis, $I_{s}$, is continuous and arrive at the following
equations: 
\begin{align}
I_{s}=w\left[\mathcal{\mathcal{J}}_{N}\left(x=0^{+}\right)+\mathcal{\mathcal{J}}_{N}\left(x=0^{-}\right)\right]\label{eq:spin_current_N}
\end{align}
\begin{align}
I_{s}=A_{J}\mathcal{\mathcal{J}}_{F}\left(z=0^{+}\right)\label{eq:spin_current_F}
\end{align}
Next, the spin current in each channel stems from the drop of electro-chemical
potential between ferromagnetic metal and 2D metal is given by $I_{I}^{\sigma}=\left(G^{\sigma}/e^{2}\right)\left[\mu_{F}^{\sigma}\left(z=0\right)-\mu_{N}^{\sigma}\left(x=0\right)\right]$.
The total spin current and charge current are thus given by:

\begin{align}
I=\sum_{\sigma}I_{I}^{\sigma}\label{eq:interface_current}
\end{align}
\begin{align}
I_{s}=\sum_{\sigma}\sigma I_{I}^{\sigma}\label{eq:interface_spin_current}
\end{align}

Finally, by solving Eqs.~\eqref{eq:inject_parallel},~\eqref{eq:spin_current_N},~\eqref{eq:spin_current_F},~\eqref{eq:interface_current},~\eqref{eq:interface_spin_current},
we arrive at the solutions of $\boldsymbol{s}\left(0\right)$, $b$,
and $I_{s}$. Then, the difference in the nonlocal resistance between
quantum axis pointing in $\vec{\hat{n}}_{p}$ and quantum axis pointing
in $-\vec{\hat{n}}_{p}$ can be evaluated by plugging the solution
of $s^{x}\left(0\right)$ and $s^{z}\left(0\right)$ into the following
equation:

\begin{align}
R_{\textrm{nl}}\left(x\right)= & \frac{wJ_{y}\left(x\right)}{I\sigma_{N}}\nonumber \\
= & \frac{wD}{I\sigma_{N}}\left[\theta_{\textrm{sH}}\partial_{x}s^{z}\left(x\right)-(\theta_{\textrm{sH}}l_{R}^{-1}+l_{\textrm{DMC}}^{-1})s^{x}\left(x\right)\right]\nonumber \\
= & R_{\textrm{nl},\textrm{sH}}+R_{\textrm{nl},\textrm{EE}}+R_{\textrm{nl},\textrm{DMC}}\label{eq:Rnl}
\end{align}

Therefore, the difference in the nonlocal resistance between quantum
axis pointing in $\vec{\hat{n}}_{p}$ and quantum axis pointing in
$-\vec{\hat{n}}_{p}$ is given by: 
\begin{align}
\Delta R_{nl}\left(x\right) & =R_{0}C_{\mathrm{inj}}e^{-\tilde{q}\cos\theta_{L}x}f(\vec{\hat{n}}_{p},\omega_{L}),
\end{align}
where the dimensionless factors $f(\vec{\hat{n}}_{p},\omega_{L})$
and $C_{\mathrm{inj}}$ read: 
\begin{widetext}
\begin{align}
f(\vec{\hat{n}_{p}},\omega_{L}) & =\left\{ \left[-\theta_{\textrm{sH}}\tilde{q}l_{s}\cos\theta_{p}-\frac{l_{s}}{l_{\textrm{DMC}}}\left(\sin\theta_{L}\cos\theta_{p}+\sin\theta_{p}\cos\varphi_{p}\cos\theta_{L}\right)\right]\cos\left[\left(l_{R}^{-1}-\tilde{q}\sin\theta_{L}\right)x\right]\right.\nonumber \\
 & \left.+\left[-\theta_{\textrm{sH}}\tilde{q}l_{s}\sin\theta_{p}\cos\varphi_{p}+\frac{l_{s}}{l_{\textrm{DMC}}}\left(\cos\theta_{L}\cos\theta_{p}-\sin\theta_{p}\cos\varphi_{p}\sin\theta_{L}\right)\right]\sin\left[\left(l_{R}^{-1}-\tilde{q}\sin\theta_{L}\right)x\right]\right\} \label{eq:f full form}
\end{align}
\begin{equation}
C_{\mathrm{inj}}=\frac{2\left(\frac{G}{G_{F}}P_{F}\frac{1-P_{J}^{2}}{1-P_{F}^{2}}+P_{J}\right)}{\frac{2G_{N}}{G_{F}}\left[1+\frac{G}{G_{F}}\frac{1-P_{J}^{2}}{1-P_{F}^{2}}\right]\tilde{q}l_{s}+\frac{G}{G_{F}}\left(1-P_{J}^{2}\right)\left[\cos\theta_{L}+\left(\tilde{q}l_{s}-\cos\theta_{L}\right)\sin^{2}\theta_{p}\sin^{2}\varphi_{p}\right]}
\end{equation}
\end{widetext}
where  $\theta_{L} = \tfrac{1}{2} \tan^{-1}\left[\omega_{L}\tau_{s}/\left(1-l_{s}^{2}/l_{R}^{2}\right)\right]$,
$\tilde{q}l_{s}=\left[\left(1-l_{s}^{2}/l_{R}^{2}\right)^{2}+\left(\omega_{L}\tau_{s}\right)^{2}\right]^{1/4}$,
$G_{F}=A_{J}\sigma_{F}/\lambda_{F}$ is the conductance of the ferromagnet,
$P_{J}=\left|G^{\uparrow}-G^{\downarrow}\right|/G$ is the interfacial
current poalrization, $P_{F}=\left(\sigma_{F}^{\uparrow}-\sigma_{F}^{\downarrow}\right)/\left(\sigma_{F}^{\uparrow}+\sigma_{F}^{\downarrow}\right)$
is the current polarization of the ferromagnetic metal, $G_{N}=w\sigma_{N}/l_{s}$
is the characteristic conductance of the 2D metal, and $G=G^{\uparrow}+G^{\downarrow}$
is the total interfacial conductance. Note that we track to all order
in the conversion factors ($\theta_{\textrm{sH}}$, $l_{s}/l_{\textrm{DMC}}$,
$l_{s}/l_{\textrm{R}}$) here and only track to the first order in
every conversion factor in the main text.

Lastly, $\Delta R_{nl}\left(x\right)$ can be decomposed into the
SHE, EE, and DMC contributions:
\begin{widetext}
\begin{align}
\Delta R_{\textrm{nl},\textrm{sH}}\left(x\right) & =\frac{2wD}{I\sigma_{N}}\theta_{\textrm{sH}}\partial_{x}s^{z}\nonumber \\
 & =R_{0}C_{\mathrm{inj}}e^{-\tilde{q}\cos\theta_{L}x}\left\{ \left[-\theta_{\textrm{sH}}\tilde{q}l_{s}\cos\theta_{p}+\frac{\theta_{\textrm{sH}}l_{s}}{l_{\textrm{R}}}\left(\sin\theta_{L}\cos\theta_{p}+\sin\theta_{p}\cos\varphi_{p}\cos\theta_{L}\right)\right]\cos\left[\left(l_{R}^{-1}-\tilde{q}\sin\theta_{L}\right)x\right]\right.\nonumber \\
 & \left.+\left[-\theta_{\textrm{sH}}\tilde{q}l_{s}\sin\theta_{p}\cos\varphi_{p}-\frac{\theta_{\textrm{sH}}l_{s}}{l_{\textrm{R}}}\left(\cos\theta_{L}\cos\theta_{p}-\sin\theta_{p}\cos\varphi_{p}\sin\theta_{L}\right)\right]\sin\left[\left(l_{R}^{-1}-\tilde{q}\sin\theta_{L}\right)x\right]\right\} \label{eq:deltaR sH}
\end{align}
\begin{align}
\Delta R_{\textrm{nl},\textrm{EE}}\left(x\right) & =-\frac{2wD}{I\sigma_{N}}\theta_{\textrm{sH}}l_{R}^{-1}s^{x}\nonumber \\
 & =R_{0}C_{\mathrm{inj}}e^{-\tilde{q}\cos\theta_{L}x}\left\{ -\frac{\theta_{\textrm{sH}}l_{s}}{l_{\textrm{R}}}\left(\sin\theta_{L}\cos\theta_{p}+\sin\theta_{p}\cos\varphi_{p}\cos\theta_{L}\right)\cos\left[\left(l_{R}^{-1}-\tilde{q}\sin\theta_{L}\right)x\right]\right.\nonumber \\
 & \left.+\frac{\theta_{\textrm{sH}}l_{s}}{l_{\textrm{R}}}\left(\cos\theta_{L}\cos\theta_{p}-\sin\theta_{p}\cos\varphi_{p}\sin\theta_{L}\right)\sin\left[\left(l_{R}^{-1}-\tilde{q}\sin\theta_{L}\right)x\right]\right\} \label{eq:deltaR Edelstein}
\end{align}
\begin{align}
\Delta R_{\textrm{nl},\textrm{DMC}}\left(x\right) & =-\frac{2wD}{I\sigma_{N}}l_{\textrm{DMC}}^{-1}s^{x}\nonumber \\
 & =R_{0}C_{\mathrm{inj}}e^{-\tilde{q}\cos\theta_{L}x}\left\{ -\frac{l_{s}}{l_{\textrm{DMC}}}\left(\sin\theta_{L}\cos\theta_{p}+\sin\theta_{p}\cos\varphi_{p}\cos\theta_{L}\right)\cos\left[\left(l_{R}^{-1}-\tilde{q}\sin\theta_{L}\right)x\right]\right.\nonumber \\
 & \left.+\frac{l_{s}}{l_{\textrm{DMC}}}\left(\cos\theta_{L}\cos\theta_{p}-\sin\theta_{p}\cos\varphi_{p}\sin\theta_{L}\right)\sin\left[\left(l_{R}^{-1}-\tilde{q}\sin\theta_{L}\right)x\right]\right\} \label{eq:deltaR DMC}
\end{align}
\end{widetext}

\section{Scattering rates and Sources}

\label{app:scattering rates}

The source term $\vec{\mathcal{Q}}(\vec{r},t)$ on the right-hand side of the 
equation for the spin density (cf. Eq.~\eqref{eq:spin_density evo}) is given by
the following expressions:
\begin{align}
\mathcal{Q}^{x}(\vec{r},t)=-\frac{s^{x}\left(\vec{r},t\right)}{\tau_{\textrm{EY}}}-2\alpha_{\textrm{asp}}\frac{J_{y}\left(\vec{r},t\right)}{v_{F}}-2\alpha_{\textrm{R}}\frac{\mathcal{J}_{x}^{z}\left(\vec{r},t\right)}{v_{F}}
\end{align}
\begin{align}
\mathcal{Q}^{y} (\vec{r},t)=-\frac{s^{y}\left(\vec{r},t\right)}{\tau_{\textrm{EY}}}+2\alpha_{\textrm{asp}}\frac{J_{x}\left(r,t\right)}{v_{F}}-2\alpha_{\textrm{R}}\frac{\mathcal{J}_{y}^{z}\left(\vec{r},t\right)}{v_{F}}
\end{align}
\begin{align}
\mathcal{Q}^{z}(\vec{r},t)=-\frac{s^{z}\left(\vec{r},t\right)}{\tau_{\textrm{EY}}^{\perp}}+2\alpha_{\textrm{R}}^{\perp}\left(\frac{\mathcal{J}_{x}^{x}\left(\vec{r},t\right)}{v_{F}}+\frac{\mathcal{J}_{y}^{y}\left(\vec{r},t\right)}{v_{F}}\right)
\end{align}

Next, the source term $\chi_{i}^{a}(\vec{r},t)$ of the time-evolution equation
of the spin density (cf. Eq.~\eqref{eq:spin_current_density evo}) is
given by the following expressions:
\begin{align}
\chi_{x}^{z}(\vec{r},t)=-\frac{\mathcal{J}_{x}^{z}\left(\vec{r},t\right)}{\tau_{c}}+\alpha_{\textrm{sk}}J_{y}\left(\vec{r},t\right)+\alpha_{\textrm{R}}v_{F}s^{x}\left(\vec{r},t\right)
\end{align}
\begin{align}
\chi_{y}^{z}(\vec{r},t)=-\frac{\mathcal{J}_{y}^{z}\left(\vec{r},t\right)}{\tau_{c}}-\alpha_{\textrm{sk}}J_{x}\left(\vec{r},t\right)+\alpha_{\textrm{R}}v_{F}s^{y}\left(\vec{r},t\right)
\end{align}
\begin{align}
\chi_{x}^{x}(\vec{r},t)=-\frac{\mathcal{J}_{x}^{x}\left(r,t\right)}{\tau_{c}^{\prime}}-\alpha_{\textrm{R}}^{\perp}v_{F}s^{z}\left(\vec{r},t\right)-\alpha_{\textrm{LD}}\mathcal{J}_{y}^{y}\left(\vec{r},t\right)
\end{align}
\begin{align}
\chi_{y}^{y}(\vec{r},t)=-\frac{\mathcal{J}_{y}^{y}\left(\vec{r},t\right)}{\tau_{c}^{\prime}}-\alpha_{\textrm{R}}^{\perp}v_{F}s^{z}\left(\vec{r},t\right)-\alpha_{\textrm{LD}}\mathcal{J}_{x}^{x}\left(\vec{r},t\right)
\end{align}
\begin{align}
\chi_{x}^{y}(\vec{r},t)=-\frac{\mathcal{J}_{x}^{y}\left(\vec{r},t\right)}{\tau_{c}^{\prime\prime}}+\alpha_{\textrm{LD}}^{\perp}\mathcal{J}_{y}^{x}\left(\vec{r}s,t\right)
\end{align}
\begin{align}
\chi_{y}^{x}(\vec{r},t)=-\frac{\mathcal{J}_{y}^{x}\left(\vec{r},t\right)}{\tau_{c}^{\prime\prime}}+\alpha_{\textrm{LD}}^{\perp}\mathcal{J}_{x}^{y}\left(\vec{r},t\right)
\end{align}
Finally, in terms of the quantum mechanical amplitudes for scattering with a single 
impurity,  the  various scattering  and relaxation rates are given by the following 
expressions:
\begin{align}
\alpha_{\textrm{asp}}=\frac{-2\pi n_{\textrm{imp}}}{\hbar}N_{0}\textrm{Re}\left(\gamma_{I}\gamma_{R}^{\star}\right)
\end{align}
\begin{align}
\alpha_{\textrm{sk}}=\frac{\pi n_{\textrm{imp}}}{\hbar}N_{0}\textrm{Im}\left(\gamma_{I}\gamma_{0}^{\star}\right)
\end{align}
\begin{align}
\alpha_{\textrm{R}}=\frac{n_{\textrm{imp}}}{\hbar}\left[\textrm{Re}\left(\gamma_{R}\right)+\pi N_{0}\textrm{Im}\left(\left(\gamma_{0}+\gamma_{I}\right)\gamma_{R}^{\star}\right)\right]
\end{align}
\begin{align}
\alpha_{\textrm{R}}^{\perp}=\frac{n_{\textrm{imp}}}{\hbar}\left[\textrm{Re}\left(\gamma_{R}\right)+\pi N_{0}\textrm{Im}\left(\left(\gamma_{0}-\gamma_{I}\right)\gamma_{R}^{\star}\right)\right]
\end{align}
\begin{align}
\frac{1}{\tau_{c}}=\frac{\pi n_{\textrm{imp}}}{2\hbar}N_{0}\left[\left|\gamma_{0}\right|^{2}+3\left|\gamma_{I}\right|^{2}+4\left|\gamma_{R}\right|^{2}\right]
\end{align}
\begin{align}
\frac{1}{\tau_{c}^{\prime}}=\frac{\pi n_{\textrm{imp}}}{2\hbar}N_{0}\left[\left|\gamma_{0}\right|^{2}+\left|\gamma_{I}\right|^{2}+6\left|\gamma_{R}\right|^{2}\right]
\end{align}
\begin{align}
\frac{1}{\tau_{c}^{\prime\prime}}=\frac{\pi n_{\textrm{imp}}}{2\hbar}N_{0}\left[\left|\gamma_{0}\right|^{2}+\left|\gamma_{I}\right|^{2}+2\left|\gamma_{R}\right|^{2}\right]
\end{align}
\begin{align}
\frac{1}{\tau_{\textrm{EY}}}=\frac{2\pi n_{\textrm{imp}}}{\hbar}N_{0}\left(\left|\gamma_{I}\right|^{2}+\left|\gamma_{R}\right|^{2}\right)
\end{align}
\begin{align}
\frac{1}{\tau_{\textrm{EY}}^{\perp}}=\frac{4\pi n_{\textrm{imp}}}{\hbar}N_{0}\left|\gamma_{R}\right|^{2}
\end{align}
\begin{align}
\alpha_{\textrm{LD}}=\frac{\pi n_{\textrm{imp}}}{\hbar}N_{0}\left[\textrm{Re}\left(\gamma_{0}\gamma_{I}^{\star}\right)+\left|\gamma_{R}\right|^{2}\right]
\end{align}
\begin{align}
\alpha_{\textrm{LD}}^{\perp}=\frac{\pi n_{\textrm{imp}}}{\hbar}N_{0}\left[\textrm{Re}\left(\gamma_{0}\gamma_{I}^{\star}\right)-\left|\gamma_{R}\right|^{2}\right]
\end{align}

\end{document}